\documentclass{aa}  

\bibpunct{(}{)}{;}{a}{}{,} 

\usepackage[english]{babel}
\usepackage[utf8]{inputenc}
\usepackage[T1]{fontenc}
\usepackage[]{tocloft}
\usepackage{lipsum}

\usepackage{amsmath}
\usepackage{amssymb}
\usepackage{amsfonts}
\usepackage{mathtools}
\usepackage{booktabs}
\usepackage{longtable}
\usepackage{stmaryrd}
\usepackage{mathrsfs}
\usepackage{multirow}
\usepackage{centernot}

\usepackage{fancyvrb}
\usepackage{graphicx}
\usepackage{floatrow}
\usepackage{subfig}
\usepackage{wrapfig}
\usepackage[usenames,dvipsnames,table]{xcolor}
\usepackage{cases}
\usepackage{multicol}
\usepackage{tikz}
\usepackage{eurosym} \def\{\euro{}} 
\usepackage{cmlgc}
\usepackage{color}
\usepackage[squaren,Gray]{SIunits}
\usepackage{empheq}
\usepackage[most]{tcolorbox}
\usepackage{booktabs}
\usepackage{textcomp} 
\usepackage{soul}

\usepackage{graphicx}
\usepackage{txfonts}
\usepackage[toc]{appendix}
\usepackage{url}
\definecolor{linkcolor}{rgb}{0,0,0}
\definecolor{linkcolorurl}{rgb}{0,0,1}
\usepackage[pdftex,colorlinks=true,
pdfstartview=FitV,
linkcolor= blue,
citecolor= blue,
urlcolor= blue,
hyperindex=true,
hyperfigures=true]
{hyperref}
\hypersetup{linktoc=page}

\begin{document}

   \title{Kinetic modeling of the electromagnetic precursor from an axisymmetric binary pulsar coalescence}
   
   \author{B. Crinquand 
          \and
          B. Cerutti
          \and
          G. Dubus
          }

   \institute{Univ. Grenoble Alpes, CNRS, IPAG, 38000 Grenoble, France\\
              \email{benjamin.crinquand@univ-grenoble-alpes.fr}
              }

   \date{Received 9 November 2018 /
         Accepted 14 December 2018}

 
  \abstract
  {The recent detection of gravitational waves associated with a binary neutron star merger revives interest in interacting pulsar magnetospheres. Current models predict that a significant amount of magnetic energy should be released prior to the merger, leading to electromagnetic precursor emission.}
  {In this paper, we revisit this problem in the light of the recent progress in kinetic modeling of pulsar magnetospheres. We limit our work to the case of aligned magnetic moments and rotation axes, and thus neglect the orbital motion.}
  {We perform global two-dimensional axisymmetric particle-in-cell simulations of two pulsar magnetospheres merging at a rate consistent with the emission of gravitational waves. Both symmetric and asymmetric systems are investigated.}
  {Simulations show a significant enhancement of magnetic dissipation within the magnetospheres as both stars get closer. Even though the magnetospheric configuration depends on the relative orientations of the pulsar spins and magnetic axes, all configurations present nearly the same radiative signature, indicating that a common dissipation mechanism is at work. The relative motion of both pulsars drives magnetic reconnection at the boundary between the two magnetospheres, leading to efficient particle acceleration and high-energy synchrotron emission. Polar-cap discharge is also strongly enhanced in asymmetric configurations, resulting in vigorous pair production and potentially additional high-energy radiation.}
  {We observe an increase in the pulsar radiative efficiency by two orders of magnitude over the last orbit before the merger exceeding the spindown power of an isolated pulsar. The expected signal is too weak to be detected at high energies even in the nearby universe. However, if a small fraction of this energy is channeled into radio waves, it could be observed as a non-repeating fast radio burst.}


   \maketitle
%

\section{Introduction}

On August 2017, the LIGO/Virgo detectors observed a gravitational wave signal that coincided with the measurement of a Short Gamma Ray Burst (SGRB) by the \emph{Fermi}/GBM instrument \citep{Gravitational_waves}. This unique joint detection allowed unambiguous association of SGRBs with binary neutron star mergers \citep{Gravitational_waves_2}.  
\smallskip

It is expected on theoretical grounds that these binary systems release a large amount of energy prior to the merger. First developed by Goldreich \& Lynden-Bell to explain decametric emission from the Io-Jupiter system \citep{Lynden_bell}, the DC model was recently adapted to other types of astrophysical systems, including binary neutron stars \citep{Piro,Lai_merger}. In this framework, the energy dissipation rate can go up to $\sim 10^{44}$ erg/s during the late stage of the inspiral, in the case of neutron stars with high magnetic fields ($\sim 10^{13}$ G). Some of this energy is then converted into observable radiation. It may partly be consumed by the creation of a plasma ``fireball'' \citep{1996ApJ...471L..95V, Lyutikov, Metzger}. The overall dissipated power predicted by these models is $\sim 10^{44-46}$ erg/s. In this case, the precursor should be observable from Earth.

The detection of an electromagnetic precursor would be very helpful to trigger multi-wavelength observations of upcoming mergers. Aside from this observational motivation, the interaction of plasma-filled magnetospheres has so far been poorly studied. In particular, it is not known whether the presence of another pulsar can enhance particle acceleration and how it affects pulsar spindown. To investigate a possible electromagnetic precursor to the merger, it is necessary to understand how particles can be accelerated, but the mechanisms at play are too complex to be analytically solved. Magnetohydrodynamic simulations were recently performed to model binary neutron star coalescences \citep{Palenzuela,Dionysopoulou}. Although this approach is appropriate to model the overall plasma structure, it fails to capture particle acceleration and cannot produce the shape of the electromagnetic outburst.

In recent years, kinetic modeling of pulsar magnetospheres using particle-in-cell (PIC) simulations has lead to significant progress in our understanding of magnetic dissipation, pair production and particle acceleration \citep{2014ApJ...785L..33P, 2014ApJ...795L..22C, Cerutti_main, 2015ApJ...801L..19P, 2015MNRAS.449.2759B, Philippov_RG, 2017A&A...607A.134C, 2018ApJ...858...81B}. In particular, they proved fit to account for the two-peaked structure of gamma-ray pulses \citep{Cerutti_light_curves, 2018ApJ...855...94P, 2018ApJ...857...44K}, and optical polarization properties of the Crab pulses \citep{2016MNRAS.463L..89C}. Spherical simulations were able to reproduce a quasi force-free magnetosphere and identify magnetic reconnection in the equatorial current sheet as the main acceleration mechanism \citep{Cerutti_main}. 

In this work we perform PIC simulations of a merging binary neutron star system in a 2D axisymmetrical cylindrical setup. To retain cylindrical symmetry, we keep both pulsars aligned with the symmetry axis, thus neglecting their relative orbital motion. It is likely that pulsar obliquity decreases with age, as can be seen on theoretical \citep{Philippov_obliquity} and observational \citep{Tauris_obliquity,Young_obliquity} grounds (see, however, \citealt{Beskin}). Therefore, it is relevant to study pulsars with aligned magnetic moments and rotation axes. Although the problem is intrinsically 3D, performing 2D simulations allows us to have a much better resolution by decreasing the numerical cost, and to gain physical insight into this complex problem more easily. We explore the mechanisms of particle acceleration resulting from magnetospheric interactions in several configurations. Furthermore, we compute the high-energy radiation lightcurve that would be produced during the inspiral phase of the merger, in order to assess the observability of the precursor. In Sec. \ref{sec:numerical}, we describe the numerical methods and the particular setup we simulate. We also report how the consistency of this new code was tested. The results are presented in Sec. \ref{sec:results}.

\section{Numerical methods} \label{sec:numerical}

\subsection{Particle-in-cell technique}

In this study, we use the relativistic PIC code \texttt{Zeltron} \citep{Zeltron,Cerutti_main}. In PIC methods, the plasma is treated as a set of charged macro-particles. During one time step three operations must be performed. First, the positions and velocities of the particles are evolved using the Boris push \citep{Birdsall}. The electromagnetic source terms are then collected on a Yee grid \citep{1966ITAP...14..302Y}, using an area weighting procedure. Finally, the fields are updated by solving discretized versions of Maxwell-Amp\`ere's and Maxwell-Faraday's equations on the grid. Given the spatial resolution, the time step is enforced by the Courant-Friedrich-Lewy condition \citep{Pic_review_cerutti}, so that the numerical scheme remains stable. This scheme does not enforce Maxwell-Gauss' equation, so the total electric charge deposited on the grid is not conserved. Small errors can accumulate and give rise to nonphysical charge densities. The electric field is therefore corrected every 25 time steps, by solving Poisson's equation: $\Delta (\delta \Phi) = -(4 \pi \rho - \vec{\nabla} \cdot \vec{E})$, where $\rho$ is the charge density, $\vec{E}_0$ the uncorrected field and $\vec{E} = \vec{E}_0 - \vec{\nabla} (\delta \Phi)$ the corrected electric field.
\smallskip

\begin{figure}[ht!]
    \centering
    \resizebox{\hsize}{!}{
    \def\svgwidth{1.0\textwidth}
    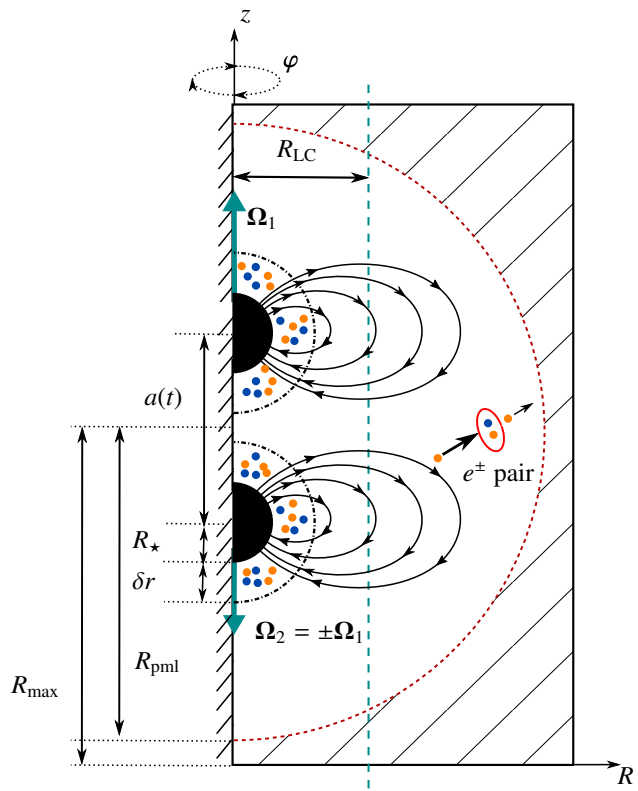 
    }
    \vspace{-0.0cm}	
    \caption{Numerical setup. The initial magnetic field lines (black solid lines) are dipolar, the magnetic moments are here identical in direction and norm. The angular velocity vectors of the two stars $\vec{\Omega}_1$ and $\vec{\Omega}_2$ are either parallel or anti-parallel, with same magnitudes. The particles are launched from the shell $R_\star \le r_i \le R_\star + \delta r$ with corotation velocity. The size of the injecting shell is exaggerated in the figure. Positrons and electrons are denoted by blue and orange disks. The dashed line marks the radius of the common light cylinder $R_{LC}=c/\Omega$. The striped zone has a resistivity profile $\lambda(r)$ (see Eq. \eqref{eq:resistivity}). The pair creation process is pictured on the right, a new pair being created at the expense of the primary particle's energy.}
    \label{fig:numerical_setup}
\end{figure}

Since we intend to model a binary pulsar, the system does no longer retain spherical symmetry, which is natural to consider for an isolated pulsar study. Instead we have implemented a 2D axisymmetric cylindrical version of the code: particles move in a 3D space but all quantities are invariant under rotations about the symmetry axis $(Oz)$. Maxwell's equations are then solved in cylindrical coordinates $(R,z)$, where $R$ is the distance from the symmetry axis. The simulated domain is defined as $(R\in [R_\mathrm{min}, R_\mathrm{max}], \ z \in [-R_\mathrm{max},R_\mathrm{max}] )$, it is centered at the origin. All simulation parameters are given in Tab. \ref{tab:parameters}. The simulated pulsars must have aligned magnetic moments and spin axes. This amounts to neglecting the orbital motion of the two pulsars, which are constantly aligned (see Fig. \ref{fig:numerical_setup}). 

\subsection{Particles} \label{sec:particles}

The numerical setup is presented in Fig. \ref{fig:numerical_setup}. The stellar radius is denoted by $R_\star$. The angular velocity vector and magnetic moment of the top (resp. the bottom) pulsar will be denoted $\vec{\Omega}_1$ and $\vec{\mathcal{M}}_1$ (resp. $\vec{\Omega}_2$ and $\vec{\mathcal{M}}_2$). They are aligned with the symmetry axis $(Oz)$. The simulation starts in complete vacuum, and the stars remain empty of particles at all times. In order to fill the magnetospheres with plasma, electron/positron ($e^\pm$) pairs are uniformly injected within thin injection shells between $R_\star$ and $R_0=R_\star+\delta r$ surrounding the stars. Charges with the right sign are then extracted from the surface, while the others rush back to the star. The particle distribution in this shell is reinitialized at every time step, in order to avoid numerical over-densities. These particles are initially corotating with the star, with toroidal velocity $v_{\varphi,i}= R \Omega_i$. Particles that leave the simulation domain are removed.
\smallskip

The density of the surface plasma is set at the Goldreich-Julian density $n_{GJ,i}=B_i \Omega_i /(2 \pi c e )$, where $B_i$ is the polar field strength at the stellar surface \citep{Goldreich_Julian}. This is consistent with the actual charge density that is extracted from the star to screen the surface electric fields. The whole electromagnetic cascade, filling the magnetosphere with $e^\pm$ plasma, is implemented in a simplified way. Following \citet{2015ApJ...801L..19P}, a critical Lorentz factor $\gamma_c$ is empirically defined as a fraction $\zeta$ of the maximal energy available to a particle, which is given by the total voltage drop $\delta V_{\mathrm{max},i} = R_0^2 B_i \Omega_i / c$ from the pole to the equator at the stellar surface. If a particle reaches a Lorentz factor\footnote{In asymmetric simulations we kept the product $B_i \Omega_i$ constant, so that the threshold $\gamma_c$ is unequivocally defined.} $\gamma = \gamma_{c,i}= \zeta e \delta V_{\mathrm{max},i} / m c^2$, and provided the local pair plasma multiplicity (defined as the ratio between the plasma density and the Goldreich-Julian density) remains smaller than $\sim 30$, an $e^\pm$ pair is created at the location of the parent particle, whose energy is consequently reduced. The resulting pair carries a fixed fraction of the parent particle's energy. The $\zeta$ factor was chosen so as to sufficiently fill the magnetosphere and reach a force-free regime \citep{Cerutti_review}. This was checked by comparing the spindown of a single pulsar to the expected values in the monopole and dipole cases. Some numerical problems arise because of the discretization of the pulsar sphere by Cartesian cells in the $(R,z)$ plane. For example, it can cause important angular inhomogeneities in the plasma injection, which impact the whole magnetosphere. This difficulty is circumvented by moderately increasing the thickness of the injecting shell and the number of macro-particles per cell.

\subsection{Electromagnetic fields} \label{sec:electromagnetic}

In most simulations (see Sec. \ref{sec:asymmetric} for asymmetric simulations with $\Omega_1 \neq \Omega_2$ and $\mathcal{M}_1 \neq \mathcal{M}_2$), $\vec{\Omega}_1$ and $\vec{\Omega}_2$ have the same magnitude, so both pulsars have the same light cylinder $R_{LC}=c/\Omega$. The two pulsars are initially separated by a distance $a_0$, and disposed symmetrically with respect to the origin. The initial magnetic field $\vec{B}_\mathrm{tot}$ is the sum of two dipole fields generated by each pulsar. As the simulation proceeds, this field is enforced inside the stars (for $r_i \leq R_0$, with $r_i$ the distance from the center of pulsar '$i$'). Pulsar spin is implemented by imposing a poloidal electric field within the pulsar (for $r_i \leq R_0$), given by
\begin{equation} \label{eq:induced_field}
    \vec{E}_i (\vec{r}) = - \dfrac{1}{c} ( \vec{\Omega}_i \times \vec{r} ) \times \vec{B}_\mathrm{tot},
\end{equation}
arising in the laboratory frame because the electric field is zero in the corotating frame, assuming that the neutron star interior is perfectly conducting. The toroidal electric field in the star is set to zero. In order to mimic an open outer boundary, a numerical resistivity $\lambda(r)$ (where $r$ is the distance from the origin) is given to the medium for $r\ge R_\mathrm{pml}$, with $R_\mathrm{pml} = 0.9 \,  R_\mathrm{max}$. This amounts to adding a resistive term to Maxwell's equations: for instance Maxwell-Faraday's equation becomes $\partial \vec{B} / \partial t = - \lambda (r) \vec{B} - c \, \vec{\nabla} \times \vec{E}$. In this part of the simulated domain the fields are critically damped. The chosen resistivity profile is chosen to be smooth, so as to avoid unwanted reflections \citep{Cerutti_main}:
\begin{equation} \label{eq:resistivity}
    \lambda (r) \propto \left( \dfrac{r - R_\mathrm{pml}}{R_\mathrm{max} - R_\mathrm{pml}} \right)^3, \quad \text{for} \ r \ge R_\mathrm{pml}.
\end{equation}
On the outer boundaries, a zero-gradient condition is enforced. Cylindrical symmetry requires the radial ($B_R,E_R$) and toroidal ($B_\varphi,E_\varphi)$ fields to be zero on the symmetry axis, whereas the vertical ($B_z,E_z$) components have no gradient on the axis. The correction potential $\delta \Phi$ is set to zero inside the pulsars, with an additional zero-gradient condition for the electric field at the stellar surfaces.
\smallskip

Initially, the electric field is zero everywhere but in the pulsar. A simulation starts with the launching of a spherical wave $(E_\theta,B_\varphi)$ distributing the electric field in the domain. If the toroidal magnetic field $B_\varphi$ is set to zero for $r\le R_0$, the simulation cannot start properly. On the other hand, if $B_\varphi$ is simply left free to evolve, some spurious $B_\varphi$ grows at a slow but constant rate inside the injection shell. This is due to a round-off error in the computation of the $\varphi$ component of $\vec{\nabla} \times \vec{E}$ in the shell ($R_\star \leq r_i \leq R_0$), which is supposed to be zero inside the star. This problem does not occur in spherical coordinates, because it is sufficient to use a single radial cell as an injection domain. In this cylindrical setup the shell must be several cells wide to enable homogeneous plasma injection. To alleviate this difficulty, a resistivity profile similar to Eq. \eqref{eq:resistivity} was given to the injection shell, where $B_\varphi$ was left free to evolve. The induced damping rate was chosen to be slightly greater than the spurious growth rate.

\newcolumntype{A}{>{\centering\arraybackslash} p{5cm}}  
\newcolumntype{D}{>{\centering\arraybackslash} p{3.2cm}}  
\newcolumntype{E}{>{\centering\arraybackslash} p{2.2cm}}  
\renewcommand{\arraystretch}{1.2}

\begin{table}[h!]
\centering
   \captionsetup{width=0.9\textwidth}
\begin{tabular}{AD}
\hline\hline
Parameter & Value \\
\hline
Number of cells &  $2000 \ (R) \times 4008 \ (z)$ \\
 Neutron star radius $R_\star/R_{LC}$ & $0.25$ \\
Injecting shell radius $R_0 /R_{LC}$ & $0.32$ \\
   $R_\mathrm{pml}/R_{LC}$ & $3.15$ \\
Skin depth $d_e / \Delta R$ & $3.5$\\
   Larmor frequency $\omega_c \Delta t$ & $8.2$ \\
Plasma frequency $\omega_p \Delta t$ &  $0.031$ \\
   Pulsar period $P /\Delta t$ & $\approx 1.2 \times 10^4$ \\
   Pair creation threshold $\zeta$ & $0.02$ \\
Start of the merger $t_0/ P$ & $2.5$ \\
  Initial number of particles per cell &  $32$ \\
  Initial inspiral speed $\dot{a}/c$ (for $a_0/R_{LC} = 1.25$) & $2.5 \times 10^{-2}$ \\
\hline
\end{tabular}
\caption{List of physical and numerical parameters used in the simulations. The plasma quantities are those evaluated at $r=R_\star$. Therefore, even if they are not resolved right at the stellar surface, the fast decrease of the fields with distance makes them well resolved a few cells away from the star.}
\label{tab:parameters}
\end{table}

\subsection{Pulsar separation}

Since the orbiting binary pulsar loses energy because of gravitational wave emission, the two stars get closer and closer. The orbital motion of the stars was not simulated, but the analytical expression for the separation $a(t)$ between them was implemented \emph{ad hoc}. For a binary star with masses $m_1$, $m_2$, separated by a distance $a$, the rate of energy loss is given by \citep{RG}:
\begin{equation}
    \dfrac{\mathrm{d}E}{\mathrm{d}t} = - \dfrac{32 G^4 (m_1 + m_2) m_1^2 m_2^2}{5 c^5 a^5}.
\end{equation}
Injecting the gravitational energy of a two-body system $E=-G m_1 m_2 /2a$ and solving the associated differential equation yields:
\begin{equation} \label{eq:separation}
    a(t)=a_0 \left( 1 - \dfrac{t}{\tau} \right)^{1/4},
\end{equation}
where $a_0$ is the initial separation. Numerical evaluation for the inspiral time $\tau$ yields
\begin{equation} \label{eq:tau}
    \tau = 0.003 \left( \dfrac{a_0}{\unit{30}{\kilo\meter}} \right)^4 \, \text{s}.
\end{equation}
The inspiral time was rescaled so as to match the code units, with $\tau \sim P (a_0 / 3 R_\star)^4$. 

The inspiral begins in the simulation only once both magnetospheres have been filled with plasma and have reached a quasi-steady state (the topology of the magnetosphere and the field strength no longer vary), which usually requires about $2.5$ rotation periods. Once this phase has elapsed, the pulsar separation decreases according to Eq. \eqref{eq:separation}. The simulation ends when the pulsars touch. Otherwise, no general relativity effect was included in the simulation.

\subsection{Modeling radiation} \label{sec:radiation}

In order to evaluate the electromagnetic signature of the merger, the radiation of accelerated particles must be computed. The procedure is described in \citet{Cerutti_light_curves}. Particles essentially cool through synchrotron radiation. The radiated power spectrum emitted by a single particle with Lorentz factor $\gamma$ moving in a perpendicular magnetic field $B_\perp$ in the ultra-relativistic limit $\gamma \gg 1$ is \citep{Blumenthal}:
\begin{equation} \label{eq:synchrotron_spectrum}
 F_\nu (\nu) = \dfrac{\sqrt{3} e^3 B_\perp}{m c^2} \dfrac{\nu}{\nu_c} \int_{\nu/\nu_c}^{+\infty} K_{5/3}(x) \, \mathrm{d}x,
\end{equation}
where $K_{5/3}$ is the modified Bessel function of order $5/3$. The critical frequency $\nu_c$ is given by
\begin{equation}
 \nu_c = \dfrac{3 e B_\perp \gamma^2}{4 \pi m c}.
\end{equation}
In the presence of an electric field, a standard procedure consists in replacing the perpendicular magnetic field $B_\perp$ by an effective field strength \citep{Kelner,Cerutti_light_curves}:
\begin{equation} \label{eq:effective_field}
 \tilde{B}_\perp = \sqrt{\left(\vec{E} + (\vec{v}/c) \times \vec{B}\right)^2 - \left(\vec{v} \cdot \vec{E} /c\right)^2}.
\end{equation}
This procedure is valid only in the ultra-relativistic limit, with $\left\lVert \vec{v} \right\rVert \approx c$, and takes into account both synchrotron and curvature radiation.  This procedure allows one to deal with the $\vec{E}\times\vec{B}$ drift. Indeed, in the wind zone, the motion of the plasma is dominated by this drift. Injecting the drift velocity $\vec{U}=c \vec{E}\times\vec{B}/B^2$ into Eq. \eqref{eq:effective_field} yields $\tilde{B}_\perp \propto \vec{E} \cdot \vec{B} \approx 0$. Physically, the plasma does not radiate much when the flow is dominated by the $\vec{E}\times\vec{B}$ drift, because in the drift frame where the electric field is zero the magnetic field strength is reduced by a factor $\Gamma \gg 1$, the Lorentz factor of the bulk flow. Keeping the actual $B_\perp$ instead of the effective field would lead to highly overestimated synchrotron losses. 
\smallskip

Moving charged particles emit ``macro-photons'' along their direction of motion. This is only valid in the ultra-relativistic limit, as the emission cone has a semi-aperture angle $1/\gamma \ll 1$. Once emitted, these photons escape the simulation (unless they hit a star). Each macro-photon conveys a power spectrum given by \eqref{eq:synchrotron_spectrum}. The output of the code is $\nu F_\nu ( \nu, R, z)$, where $F_\nu = h \nu \, \mathrm{d} N / \mathrm{d} \nu$ is the locally emitted power spectrum. In order to find the high-energy emission sites, the code outputs $\int_{\nu_0}^{+\infty} \nu F_\nu \, \mathrm{d} \ln \, \nu$, where $\nu_0 = q B_0/mc$ is a typical synchrotron frequency. In this 2D axisymmetric setup the spherical surface of a pulsar must be discretized in Cartesian coordinates (see Sects \ref{sec:particles} and \ref{sec:electromagnetic}). Right at the stellar surface, fields suffer discontinuities that produce spurious particle acceleration. Consequently, we choose to discard macro-photons emitted within $5 \%$ of the stellar surfaces. Similarly, particles that would undergo nonphysically high accelerations near the surfaces are also removed from the simulation.
\smallskip

The emitted macro-photons are collected at infinity, making an angle $\alpha_\mathrm{obs}$ with the symmetry axis. The propagation time of the photons depends on the location of its emitting particle and its direction of motion. The time delay $\delta t$ between a photon emitted at polar coordinates $(\rho,\theta)$ in the $(R,z)$ plane, propagating with an angle $\alpha$ with respect to the symmetry axis, and one emitted at the origin in the same direction, is given by:
\begin{equation}
    \delta t =\dfrac{\rho}{c} \cos( \alpha - \theta).
\end{equation}
All photons emitted at a given time step are distributed on a grid according to their time delay. Since the merger is inherently unsteady, we collect photons every 10 time steps.

\subsection{Consistency checks}

Since PIC simulations of pulsar magnetospheres had not yet been conducted in cylindrical coordinates, we checked that the output of the code against previous results. Some analytical solutions are known for an isolated pulsar magnetosphere. First, the electromagnetic solver alone was tested in vacuum. For a monopolar field $\vec{B}= B_0 (R_0/r)^2 \vec{e}_r$, the electric field can be readily computed in spherical coordinates $(r,\theta,\varphi)$ in the whole space \citep{Cerutti_review,Jackson}, using the boundary condition at $r=R_0$ given by Eq. \eqref{eq:induced_field}:
\begin{align}
\left\{
    \begin{array}{lll}
   E_r = - \dfrac{2 B_0 R_0^4}{R_{LC}} \dfrac{\cos \theta}{r^3},\\[2ex]
   E_\theta = - \dfrac{B_0 R_0^4}{R_{LC}} \dfrac{\sin \theta}{r^3}. \label{eq:test_electric_field}
 \end{array}
\right.
\end{align}
The simulated electric field of an empty magnetosphere was found to be consistent with Eq. \eqref{eq:test_electric_field}, the angular and radial dependence agreeing with this theoretical prediction. The toroidal field remained zero in the steady state. The code was then tested on a monopole force-free magnetosphere, for which an analytical solution also exists \citep{Michel_1,Michel_2,Henriksen}. In the presence of plasma a toroidal field develops, given by
\begin{equation}
    B_\varphi = - B_0 \dfrac{R_0}{R_{LC}} \dfrac{R_0}{r} \sin \theta.
\end{equation}
We recovered this behaviour with good accuracy. Finally, the single force-free aligned dipole configuration was also simulated. The spindown power of the dipole pulsar $L_0$ can be analytically estimated \citep{Cerutti_review} to scale like
\begin{equation} \label{eq:spindown_dipole}
  L_0 = \dfrac{c B_0^2 R_\star^6}{R_{LC}^2 }, 
\end{equation}
where $B_0$ is the equatorial surface field of the pulsar. This served as another quantitative test for the 2D-axisymmetric code. Previous 2D spherical simulations \citep{Cerutti_main, 2015MNRAS.449.2759B} showed that the outgoing Poynting flux indeed scales like $L_0$, provided the force-free regime is achieved. If the pair creation parameters are properly adjusted, so as to reach a force-free regime, we confirm this result in the axisymmetric setup. The quantity $L_0$ provides a useful reference to convert energy flux from code units to physical units.

\section{Results} \label{sec:results}

Aside from Sec. \ref{sec:asymmetric}, both pulsars have equal magnetic field strengths and rotation periods in the following. We first present simulations that have reached a quasi-steady state, in order to study the magnetospheric interactions at play. In Secs. \ref{sec:inspiral} and \ref{sec:asymmetric} we then study the inspiral phase and the resulting electromagnetic outburst.

\begin{figure*}[ht!]
    \centering
    \captionsetup[subfigure]{position=top, labelfont=bf,textfont=normalfont,singlelinecheck=off,justification=raggedright}
	\makebox[\textwidth][c]{
	\sidesubfloat[]{\includegraphics[width=9.0cm]{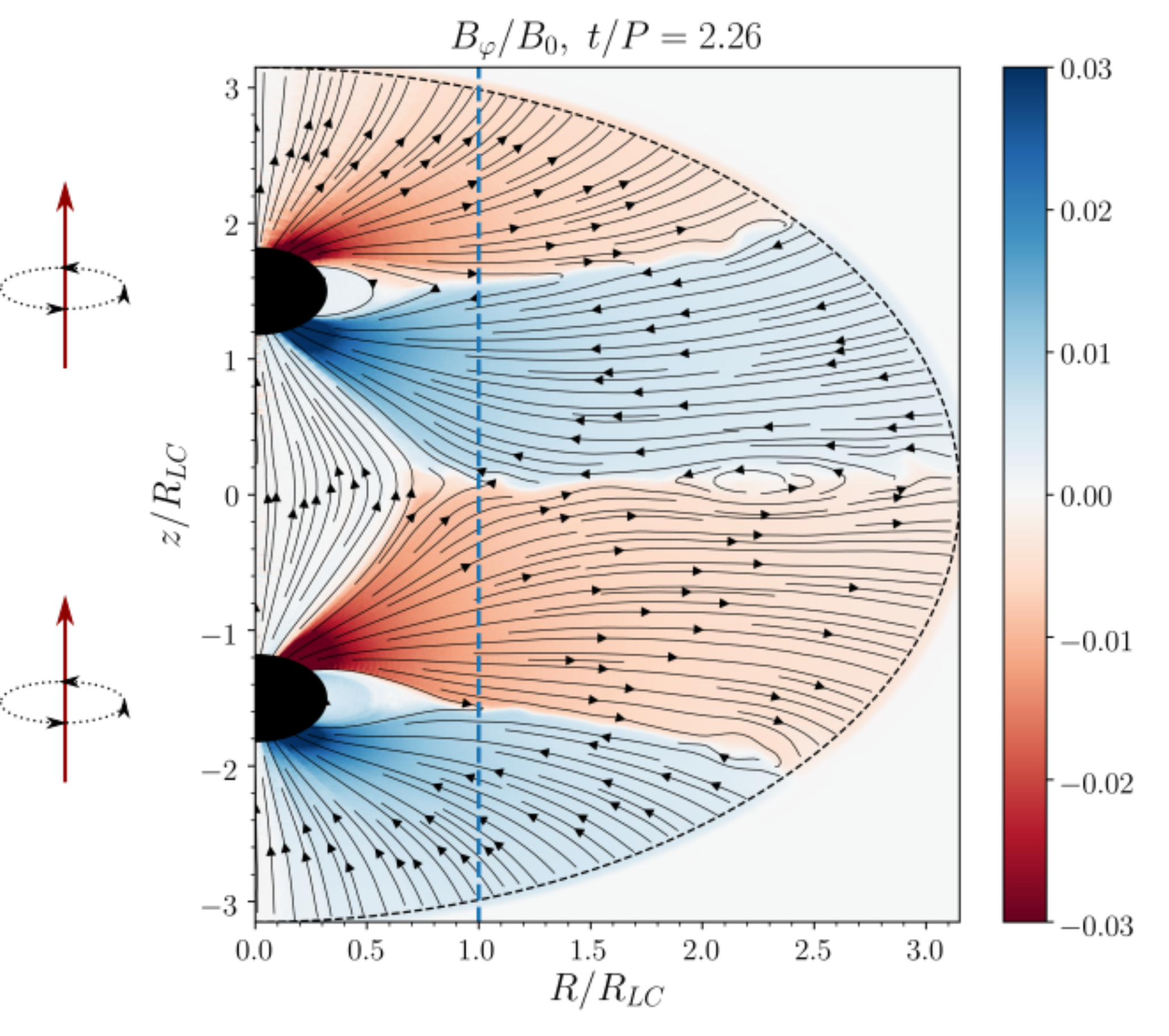}\label{fig:Bth_same_aligned}} 
	\sidesubfloat[]{\includegraphics[width=8.0cm]{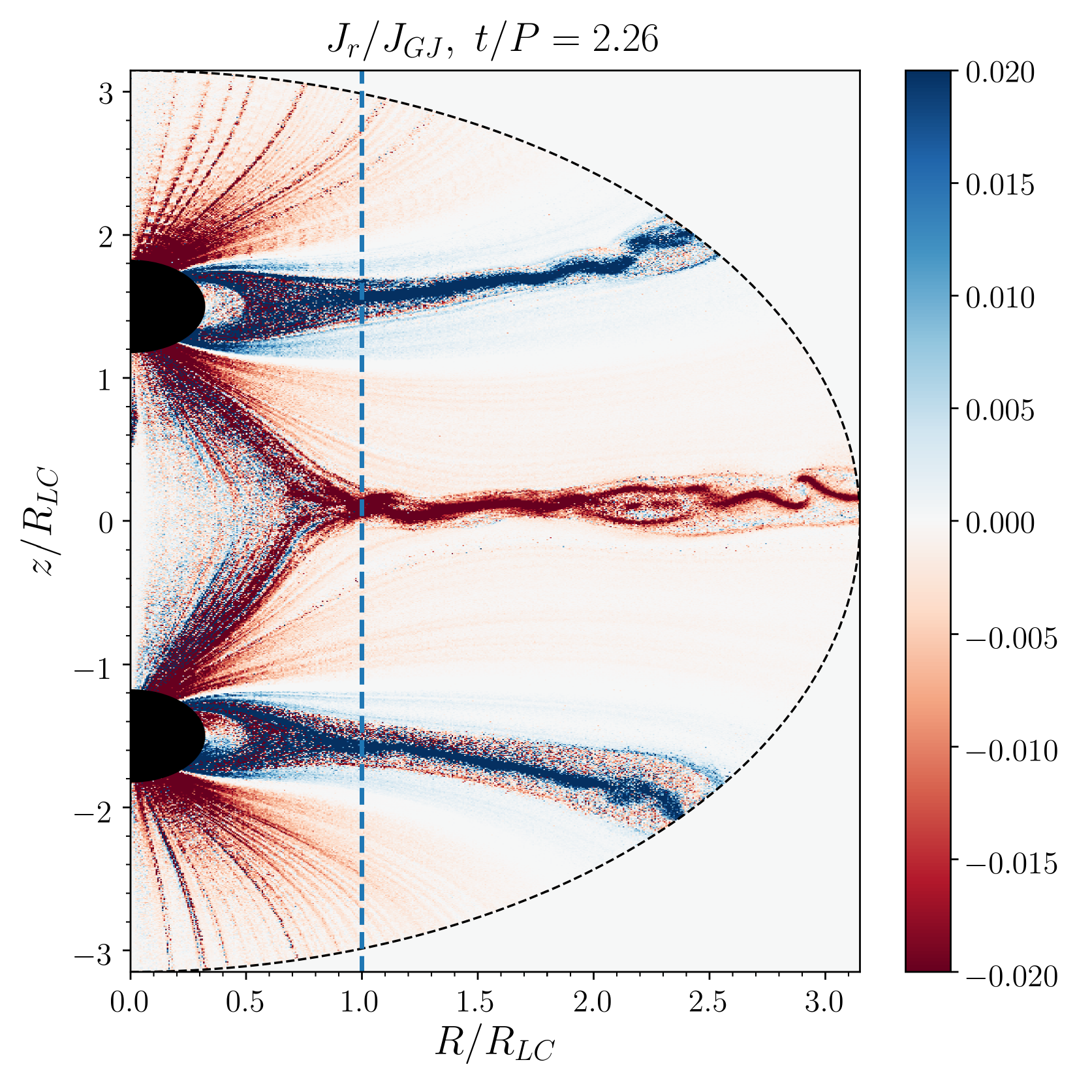}\label{fig:JR_same_aligned}} 
	}
	\caption{\textbf{(a)} Normalized toroidal magnetic field $B_\varphi /B_0$ in the parallel configuration. The poloidal magnetic field lines are depicted as black solid lines. A large magnetic island is visible in the inter-pulsar current sheet on the right. It can be recognized by its closes magnetic field lines. \textbf{(b)} Normalized radial current density $J_R / c \rho_{GJ}$ in the parallel configuration. The stellar separation is fixed at $a=3 R_{LC}$.}
\end{figure*}

\begin{figure*}[ht!]
    \centering
    \captionsetup[subfigure]{position=top, labelfont=bf,textfont=normalfont,singlelinecheck=off,justification=raggedright}
	\makebox[\textwidth][c]{
	\sidesubfloat[]{\includegraphics[width=9.0cm]{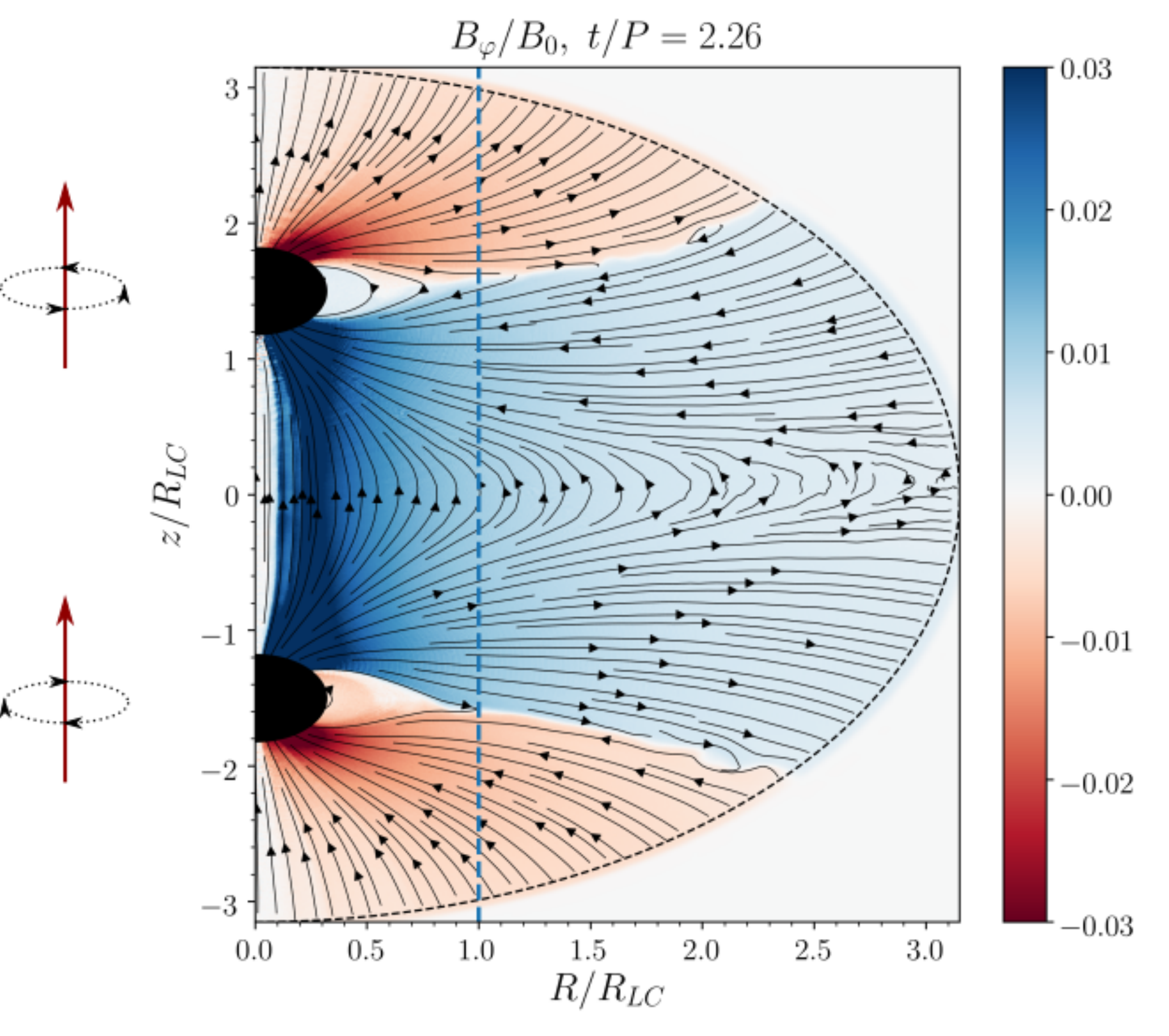}\label{fig:Bth_same_anti}} 
	\sidesubfloat[]{\includegraphics[width=8.0cm]{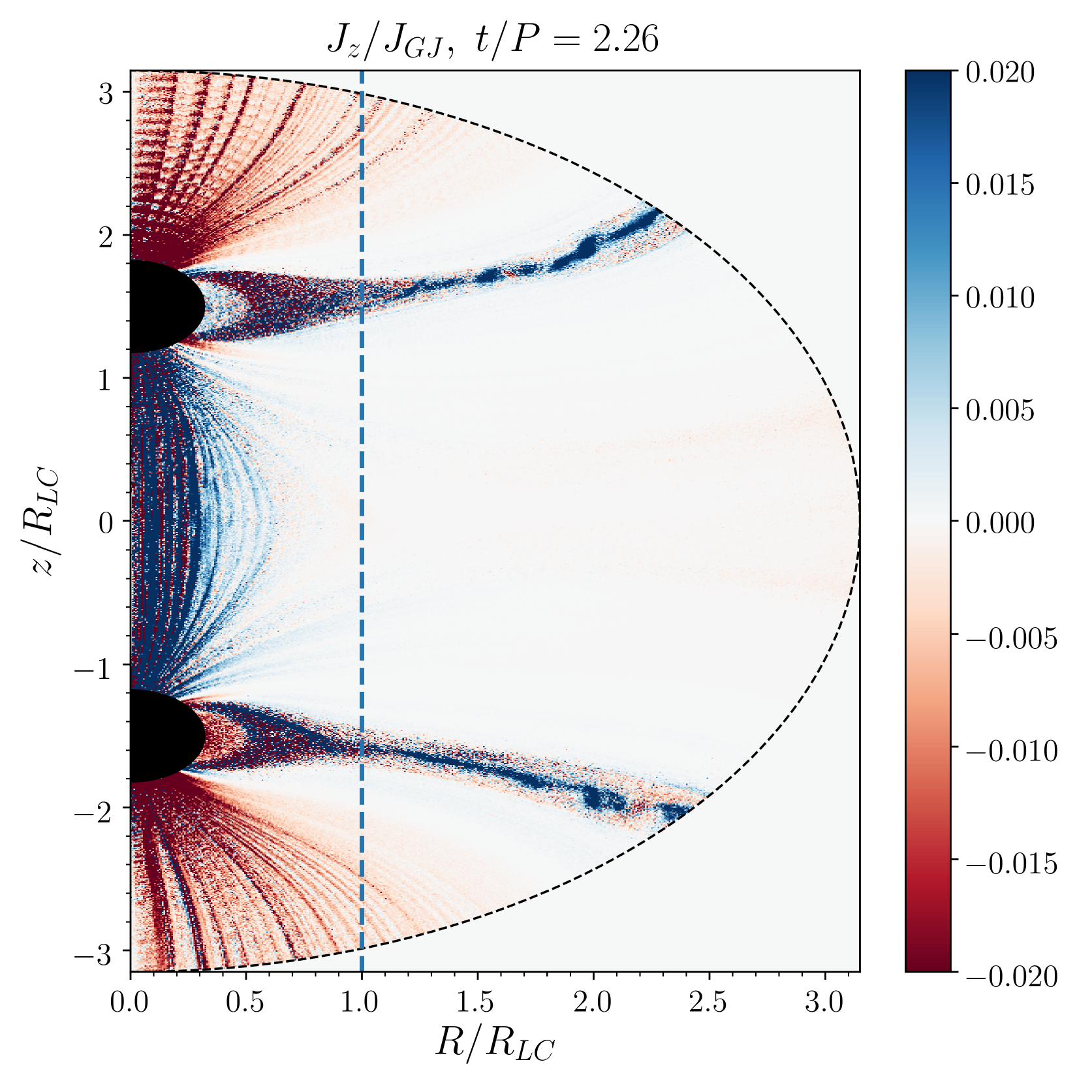}\label{fig:Jz_same_anti}} 
	}
	\caption{\textbf{(a)} Normalized toroidal magnetic field $B_\varphi /B_0$ in the anti-parallel configuration. The poloidal magnetic field lines are depicted as black solid lines. \textbf{(b)} Normalized vertical current density $J_z / c \rho_{GJ}$ in the anti-parallel configuration. The stellar separation is fixed at $a=3 R_{LC}$.}
\end{figure*}

\subsection{Magnetospheric structure} \label{sec:magnetospheric}

Eight configurations can be simulated with this axisymmetric setup, depending on the relative orientations of the magnetic moments and the rotation axes. We focus on the four configurations where the pulsars have their magnetic moments parallel or anti-parallel, and their rotation axes parallel or anti-parallel. The top pulsar has $\mathcal{M}=\vec{\mathcal{M}}\cdot \vec{e}_z >0$ and $\Omega >0$, whereas the bottom pulsar can have its magnetic moment or rotation axis reversed. Until Sec. \ref{sec:asymmetric} we only study the two configurations with the magnetic moments aligned for both stars. Indeed, symmetric simulations with opposed magnetic moments are qualitatively similar, regardless of the relative spin orientation (this is not shown in the symmetric case, but see Sec. \ref{sec:asymmetric}). The two opposed dipolar fields exclude one another, so no field line links the pulsars, and no reconnection layer occurs in between. 

\subsubsection{Parallel configuration}

\begin{figure}[ht!]
    \resizebox{\hsize}{!}{\includegraphics{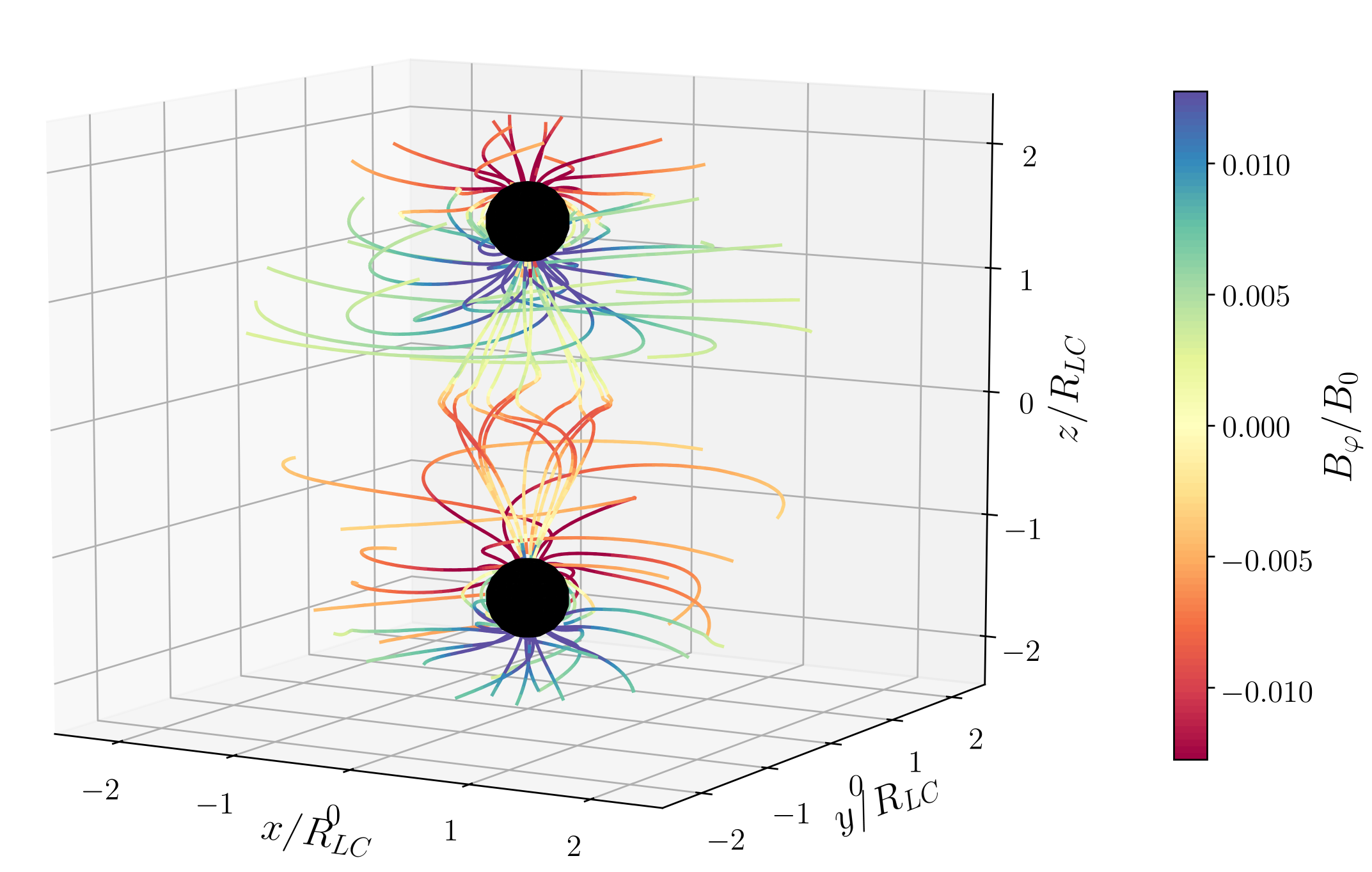}}
    \caption{Magnetic field lines in the parallel configuration. The stellar separation is fixed at $a=3 \, R_{LC}$. The pulsars are depicted as two black spheres.}
    \label{fig:field_lines_same_aligned}
\end{figure}

Fig. \ref{fig:Bth_same_aligned} shows the steady state toroidal magnetic field. Since the two pulsars have no relative motion before the merger is initiated, a large zone between them remains closed, with no outflowing currents and no toroidal fields. The main feature of this configuration is the equatorial sheet, that supports a discontinuity in the toroidal (like in the isolated pulsar case) and poloidal field. A discontinuity in the toroidal field induces a radial current sheet, visible in Fig. \ref{fig:JR_same_aligned}. Note that this ``inter-pulsar'' current sheet qualitatively differs from the ``proper'' current sheets. Indeed, it carries a current with the opposite sign, which means electrons undergo greater acceleration than with a single pulsar. It is a third promising site for particle acceleration, apart from the two proper current sheets. The location of the Y-point of this current sheet is determined exclusively by the separation between the pulsars, rather than the light cylinder. This means that unlike the isolated pulsar configuration, here reconnection occurs well within the light cylinder where the fields are much stronger, although this is more obvious at closer separations. The field lines are represented in three dimensions in Fig. \ref{fig:field_lines_same_aligned}.
\smallskip

\begin{figure*}[ht!]
    \centering
    \captionsetup[subfigure]{position=top, labelfont=bf,textfont=normalfont,singlelinecheck=off,justification=raggedright}
	\makebox[\textwidth][c]{
	\sidesubfloat[]{\includegraphics[width=9.0cm]{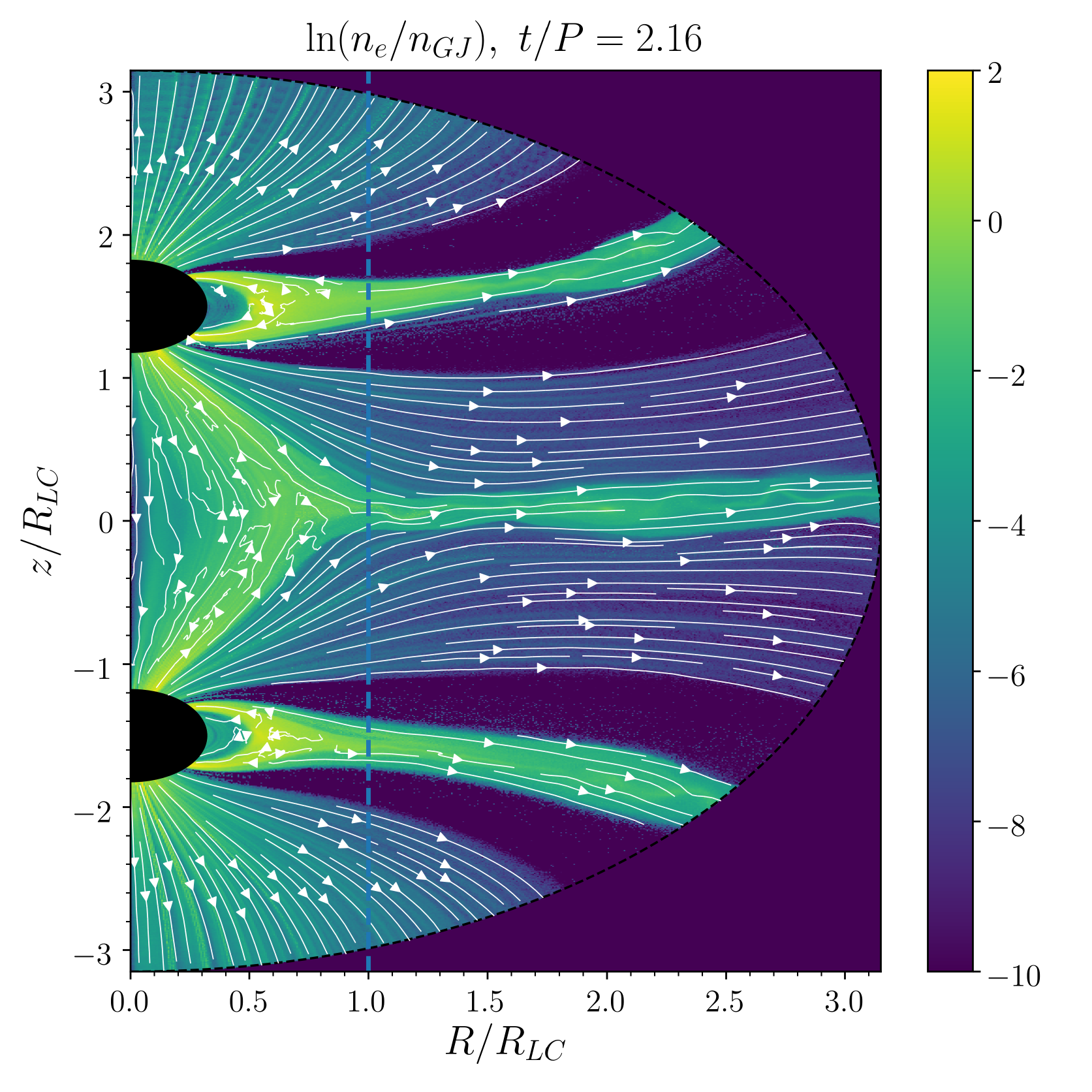}\label{fig:ne_same_aligned}} 
	\sidesubfloat[]{\includegraphics[width=9.0cm]{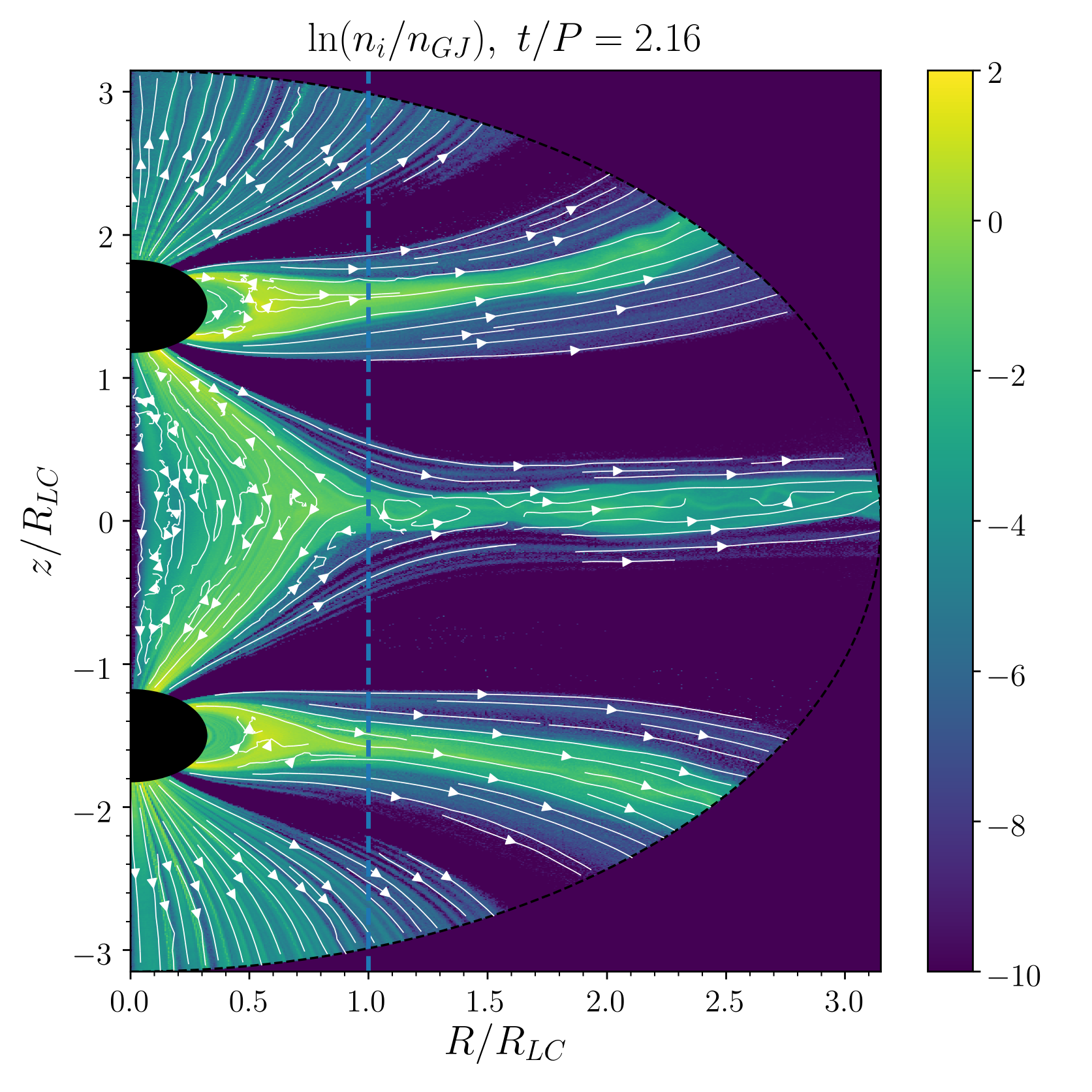}\label{fig:ni_same_aligned}} 
	}
	\caption{Logarithm of the electronic \textbf{(a)} and positronic \textbf{(b)} densities, normalized by the Goldreich-Julian density, in the parallel configuration. The bulk velocity field lines of the electrons (on the left) and positrons (on the right) are displayed in white. The stellar separation is fixed at $a=3 \, R_{LC}$.}
    \label{fig:densities_same_aligned}
\end{figure*}

Densities for the two species are shown in Fig. \ref{fig:densities_same_aligned}, as well as the bulk velocity field lines. The small discrepancies between the electronic and positronic densities imply that the plasma is globally quasi-neutral, except in the closed zones and within the current sheets. Some density gaps are visible surrounding the current sheets, as in \citet{2015ApJ...801L..19P}. In the wind zone, the solution resembles Michel's split monopole \citep{Michel_1}. The bulk velocity field is similar for both species in the wind zone, where it is approximately radial, as prescribed by the $\vec{E}\times \vec{B}$ drift velocity of the particles. A difference can be found in the inter-pulsar current sheet for instance: electrons flow outward whereas positrons move back towards the pulsars.

\subsubsection{Anti-parallel configuration} 

\begin{figure}[ht!]
    \resizebox{\hsize}{!}{\includegraphics{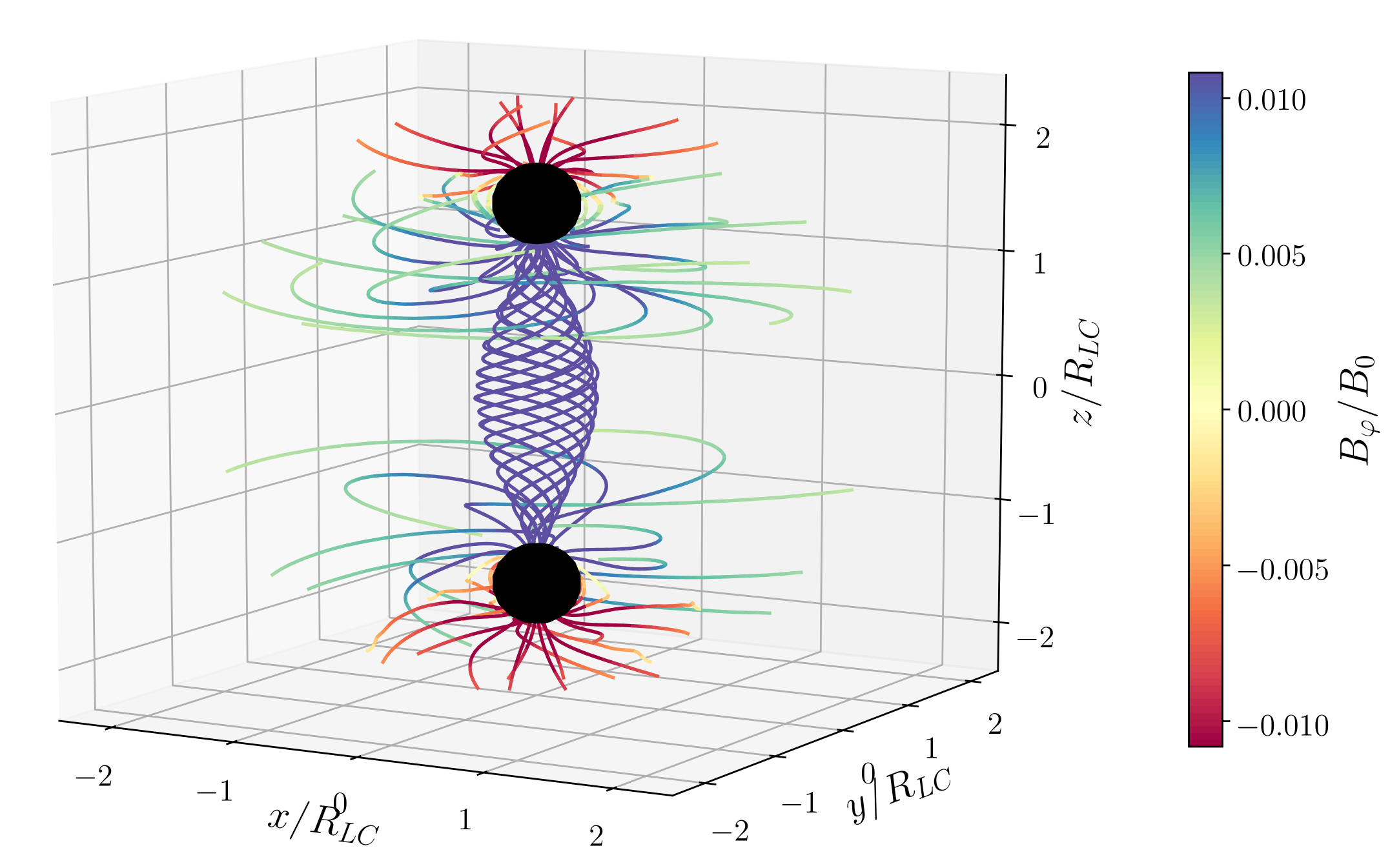}}
    \caption{Magnetic field lines in the anti-parallel configuration. The stellar separation is fixed at $a=3 \, R_{LC}$. The pulsars are depicted as two black spheres.}
    \label{fig:field_lines_same_anti}
\end{figure}

The initial magnetic field is the same as in the parallel case, but here the two stars spin in opposite directions. Consequently, the field lines linking the two stars are twisted (see Fig. \ref{fig:field_lines_same_anti}), while a strong toroidal field develops in between, as can be seen in Fig. \ref{fig:Bth_same_anti}. We can relate this configuration to the DC model \citep{Lai_merger,Piro} which considers a magnetic neutron star orbiting a non-magnetic but perfectly conducting companion. The motion of the non-magnetic star with respect to the magnetic field of the primary induces an electromotive force, which in turn drives currents along the magnetic field lines emerging from the magnetic star. The energy is dissipated because of the plasma and stellar resistivity. This model allows us to shed light on this magnetic configuration: the toroidal field between the pulsars induces a strong current flowing from the down pulsar to the top, as can be seen in Fig. \ref{fig:Jz_same_anti}. No instability prevents $B_\varphi$ from growing in these 2D simulations.
\smallskip

In both configurations, the single pulsar current sheets seem to be repelled from the midway equator, although they both carry opposite electric charges. This is a magnetic pressure effect. The presence of another pulsar generates a stronger toroidal field outside the light cylinder, which induces a magnetic pressure acting perpendicular to the field lines. The equilibrium position of the current sheets results from a balance between magnetic and kinetic plasma pressure. Close to the outer boundary, the current sheets become unstable due to the kink instability. It is also subject to the tearing instability. Magnetic islands are generated at the Y-point, then advected away (see Fig. \ref{fig:Bth_same_aligned}). 

The latitudinal stripes visible in Figs. \ref{fig:JR_same_aligned} and \ref{fig:Jz_same_anti} likely stem from the discretization of the sphere, as mentioned in Sec. \ref{sec:particles}, and should not be considered physical. However, the radial current oscillations have a physical basis. If some unscreened parallel electric field lies at the stellar surface, particles are accelerated and induce pair creation. Then the local plasma density rises, so the electric field is momentarily screened and particle acceleration is quenched. As the plasma flows away, the same process starts again, giving rise to these oscillations.

\subsection{Spindown} \label{sec:spindown}

The interaction between the two pulsars significantly impacts their spindown. Two steady state simulations were performed, with $a=3 \, R_{LC}$ and $a=1.25 \, R_{LC}$. We compute the Poynting energy flux flowing out of each pulsar at $r_i = 1.1 \, R_0$ separately, then we sum the results to get the total luminosity $L_\mathrm{\star}$. This is compared to the Poynting flux flowing out of the simulation $L_\mathrm{out}$, through a sphere of radius $R_\mathrm{pml}$ centered on the origin. The dissipation rate is then computed as $\epsilon =(L_\mathrm{\star}-L_\mathrm{out}) /L_\mathrm{\star}$. The results are shown in Tab. \ref{tab:poynting}. The spindown values lie between $2 \, L_0$ and $4 \, L_0$, $L_0$ being the spindown of an isolated pulsar \eqref{eq:spindown_dipole}. If the separation $a$ tends to infinity, the spindown will be $2 L_0$. If $a$ tends to $0$, the magnetic moment of the star will double and its spindown will quadruple. For intermediate separations, one pulsar can intercept outgoing field lines from the other, thus reducing its Poynting flux.
\smallskip

\newcolumntype{C}{>{\centering\arraybackslash} p{2.4cm}}  
\newcolumntype{B}{>{\centering\arraybackslash} p{2.0cm}}  

\begin{table}[ht!]
\centering
\begin{tabular}{BCC}
\hline\hline 
Separation & Parallel & Anti-parallel \\
\hline
 & $L_\mathrm{\star}/L_0=2.8$ & $L_\mathrm{\star}/L_0=3.1$ \\
 & $L_\mathrm{out}/L_0=2.1$ & $L_\mathrm{out}/L_0=2.3$ \\
\multirow{-3}{*}{$3 \, R_{LC}$} & $\epsilon=27 \%$ & $\epsilon =25 \%$ \\
\hline
 \multirow{3}{*}{$1.25 \, R_{LC}$} & $L_\mathrm{\star}/L_0=2.5$ & $L_\mathrm{\star}/L_0=3.9$ \\
 & $L_\mathrm{out}/L_0=1.5$ & $L_\mathrm{out}/L_0=2.7$ \\
 & $\epsilon=40 \%$ & $\epsilon=30 \%$ \\
\hline
\end{tabular}
\vspace*{-0.2cm}
\caption{Ingoing and outgoing energy fluxes, normalized by the luminosity of an isolated pulsar $L_0$, for both configurations at two different separations. The luminosities were computed in steady state.}
\label{tab:poynting}
\end{table}

At fixed separation, the anti-parallel total spindown is higher than the parallel spindown. The spindown does not increase when $a$ decreases in the parallel case because the presence of another pulsar does not significantly affect the magnitude of the toroidal field (since they have opposite polarities), and the Poynting flux approximately scales like $B_\varphi^2$. More field lines outgoing from one pulsar are captured by the other, diminishing the amount of energy extracted along the open field lines. On the other hand, in the anti-parallel configuration the toroidal field is amplified near the stars, which leads to a higher Poynting flux. 
\smallskip

The fraction of dissipated spindown increases as the separation decreases. It is greater in the parallel configuration. This fact seems to support the idea that the inter-pulsar current sheet is a prominent site for dissipation and particle acceleration. The absence of this layer in the anti-parallel configuration could imply that the DC mechanism is a less efficient dissipation mechanism than magnetic reconnection. However, the amplification of the toroidal field in this configuration compensates for this effect.

\subsection{Particle acceleration and high-energy radiation} \label{sec:acceleration}

The sites of pair creation can be investigated through the local quantity $\alpha (\theta)=J_r (R_0,\theta) / c \rho_{GJ}$, where $J_r$ is the (spherically) radial current flowing out of a pulsar, $\theta$ the polar angle defined with respect to this pulsar, and $\rho_{GJ}=e n_{GJ}$ the Goldreich-Julian charge density. If $1/\alpha < 1$, the charge-separated current escaping from the star fails to sustain the current required by the current sheet, so a parallel electric field develops which accelerates particles and produces pairs \citep{Beloborodov}. Pair creation must occur to supply current. This diagnostic was computed for both the up and down pulsars, following \citet{Parfrey_alpha_plot}.
\smallskip

\begin{figure}[ht!]
    \centering
    \captionsetup[subfigure]{position=top, labelfont=bf,textfont=normalfont,singlelinecheck=off,justification=raggedright}
    	\makebox[\textwidth][c]{
	\sidesubfloat[]{\includegraphics[width=1.0\textwidth]{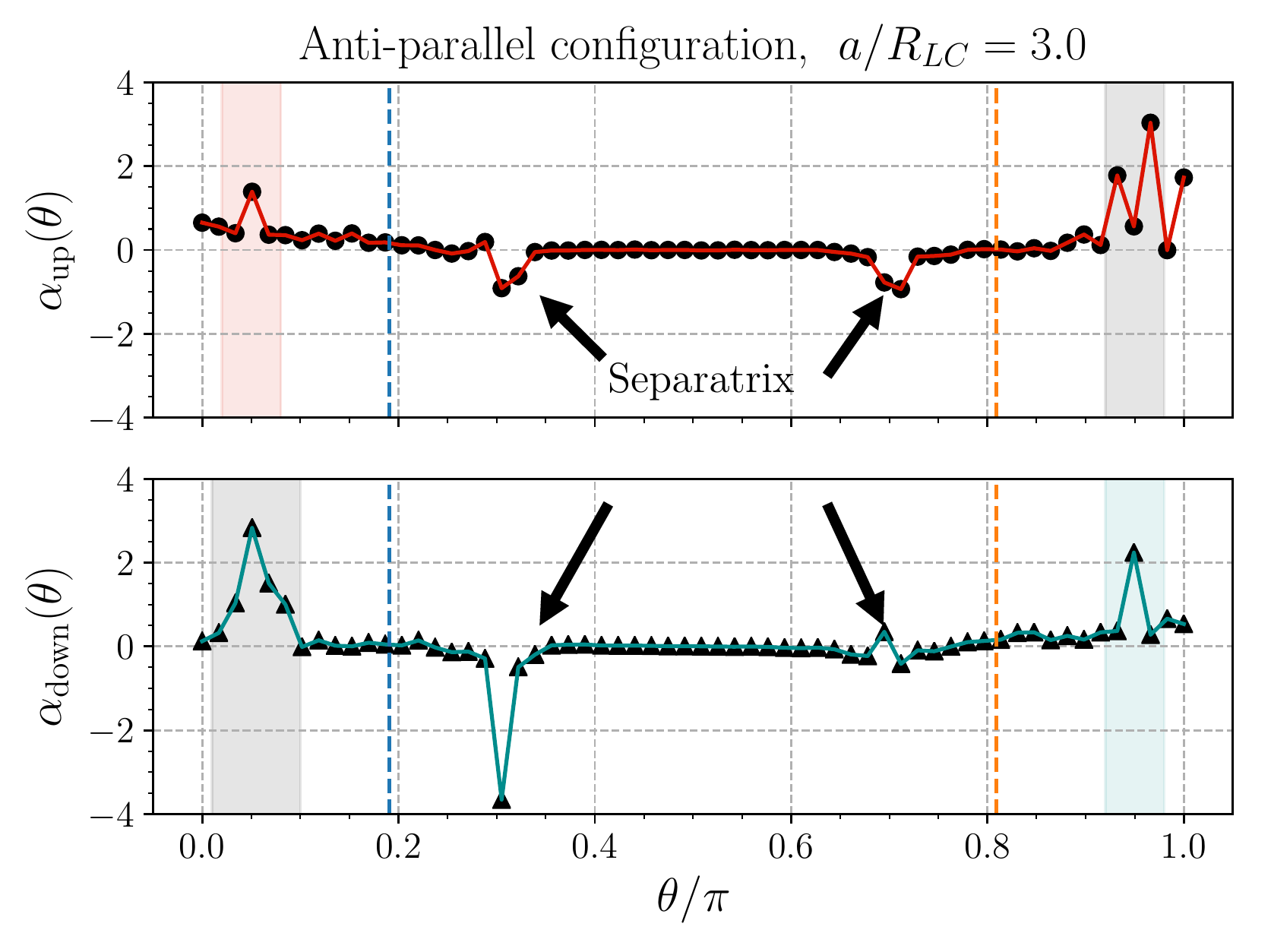}\label{fig:alpha_same_anti}} 
	}
    	\makebox[\textwidth][c]{
	\sidesubfloat[]{\includegraphics[width=1.0\textwidth]{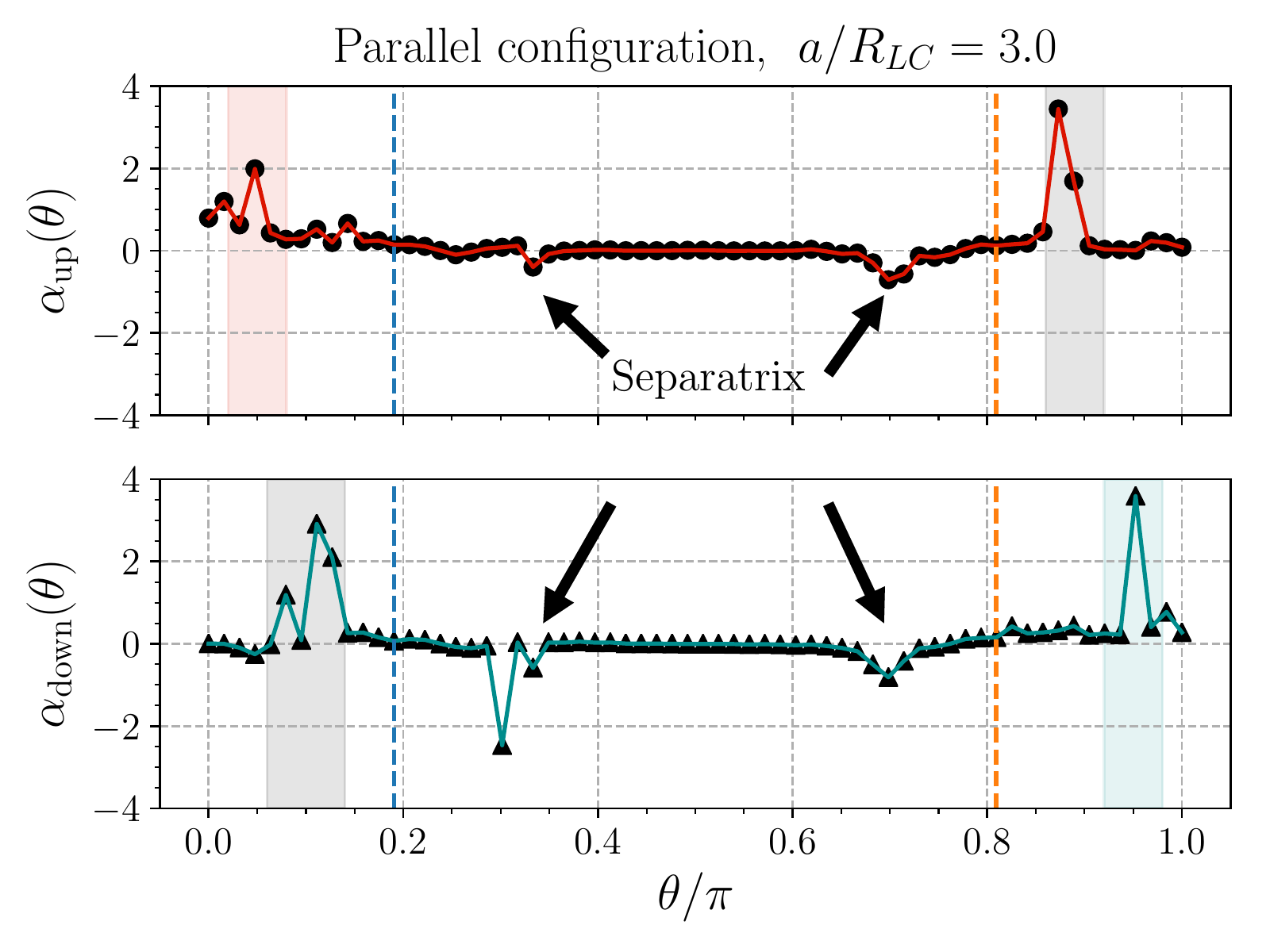}\label{fig:alpha_same_aligned}}
	}
	\caption{Plots of $\alpha(\theta)=J_r (R_0, \theta) / c \rho_{GJ}$ in the parallel \textbf{(a)} and anti-parallel \textbf{(b)} configurations, for the top (top panel) and bottom (bottom panel) pulsars. The stellar distance is fixed at $a=3 \, R_{LC}$. The dashed lines mark the locations of $\theta_{pc}$ and $\pi - \theta_{pc}$ (see Eq. \eqref{eq:polar_cap}). The grey-shaded zones indicate the inward pole regions of emission, the color-shaded zones mark the outward pole emission.}
\end{figure}

The anti-parallel configuration is presented in Fig. \ref{fig:alpha_same_anti}. The presence of the second pulsar greatly increases $\alpha$, which can reach values from $3$ to $4$ at the south pole of the top pulsar and at the north pole of the bottom pulsar (the ``inward'' poles). This is consistent with the DC model: the electromotive force between the stars drives a large current flowing from the down to the up pulsar. Pair creation is extremely efficient in between the pulsars, hence the formation of a dense pair plasma. However, this plasma remains highly magnetized and does not radiate much. At closer separation, the locations of the peaks in $\alpha$ are almost unchanged, whereas the magnitude of the inward pole peaks rises a lot. Fig. \ref{fig:alpha_same_aligned} shows the same diagnostic for the parallel configuration. Here, the magnitudes of the inward pole peaks and outward pole peaks are similar. At closer separation, the inward pole peaks move towards the stellar equators.
\smallskip

In both cases, pair creation seems to be slightly enhanced at the outward poles with respect to the isolated pulsar, $\alpha$ remaining close to $1$. Indeed, this is why periodic oscillations in current density can occur at the poles in Figs. \ref{fig:JR_same_aligned} and \ref{fig:Jz_same_anti}. We also find that the closed zone gets smaller and the polar cap larger. This is proven by the fact that the return current sheets ($\alpha <0$ marks the location of the separatrix, separating closed field lines from open ones) are located closer to the equator and further from the theoretical polar cap angles $\theta_\mathrm{pc}$ and $\pi - \theta_\mathrm{pc}$, where $\theta_\mathrm{pc}$ is given by
\begin{equation} \label{eq:polar_cap}
    \sin^2 \theta_\mathrm{pc} \approx \dfrac{R_0}{R_{LC}}.
\end{equation}
This likely results from the magnetic pressure exerted by the other pulsar. This effect is more significant in asymmetric simulations (see Sec. \ref{sec:asymmetric}).
\smallskip

\begin{figure*}[ht!]
    \centering
    \captionsetup[subfigure]{position=top, labelfont=bf,textfont=normalfont,singlelinecheck=off,justification=raggedright}
	\makebox[\textwidth][c]{
	\sidesubfloat[]{\includegraphics[width=8.0cm]{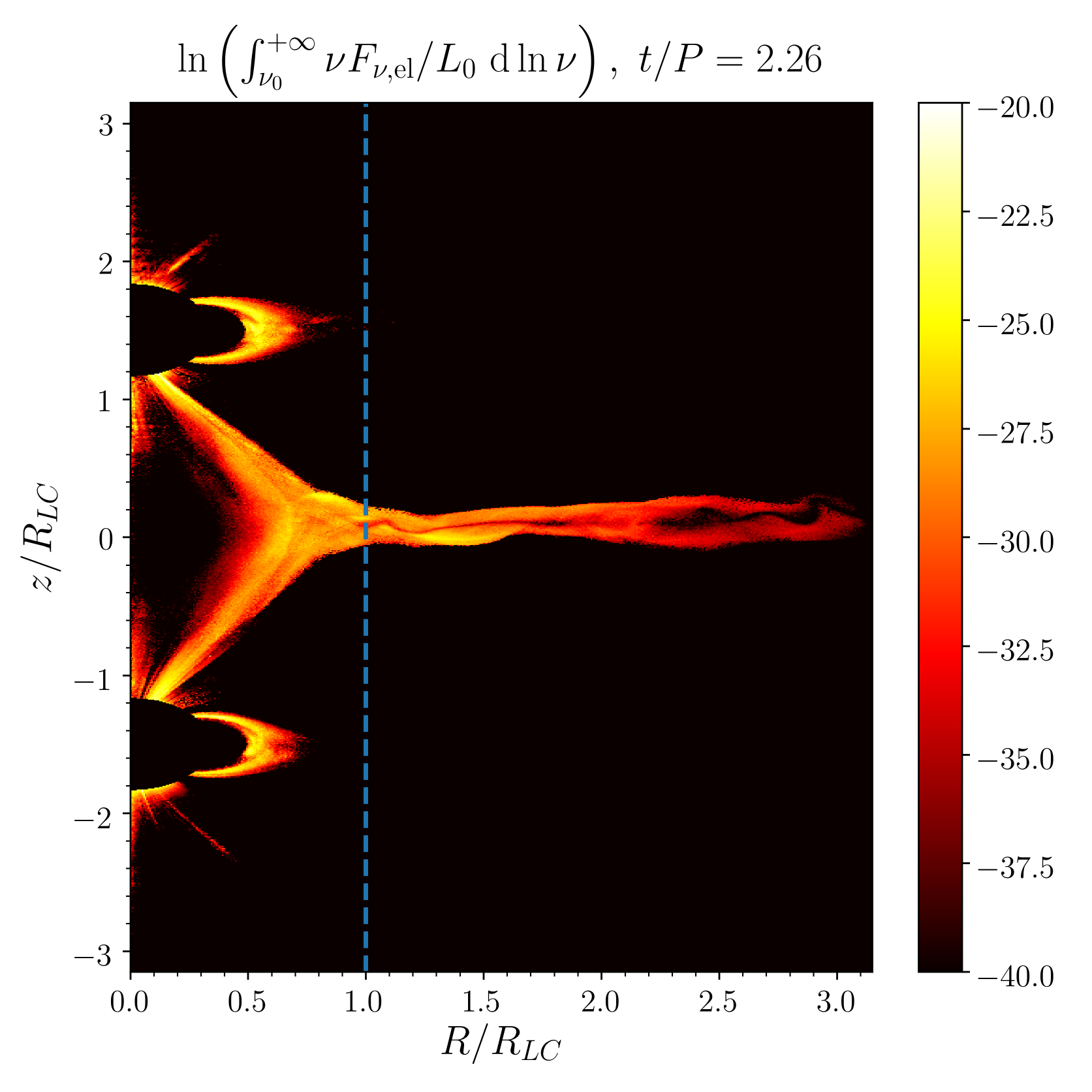}\label{fig:rad_el_same_aligned}} 
	\sidesubfloat[]{\includegraphics[width=8.0cm]{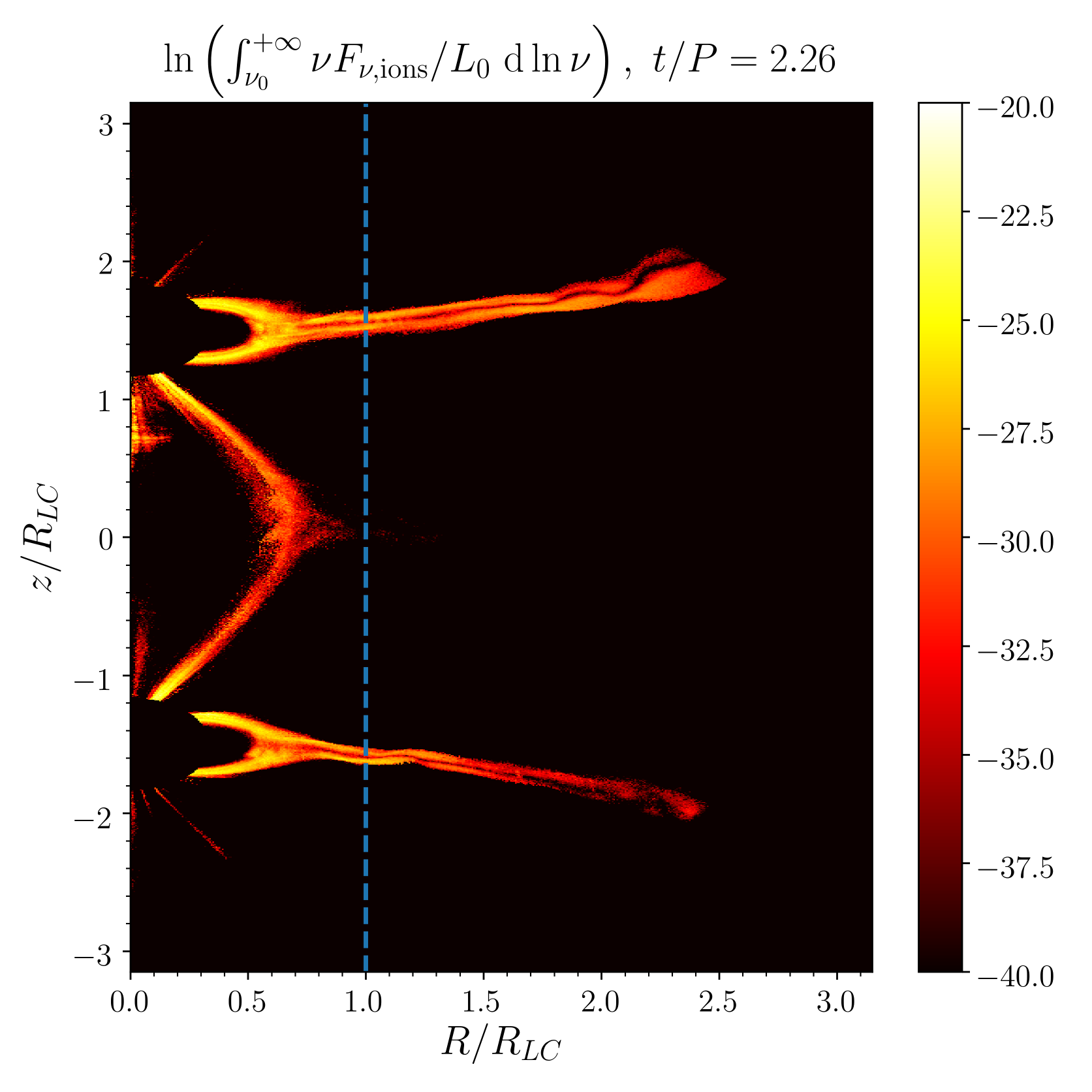}\label{fig:rad_ions_same_aligned}} 
	}
	\makebox[\textwidth][c]{
	\sidesubfloat[]{\includegraphics[width=8.0cm]{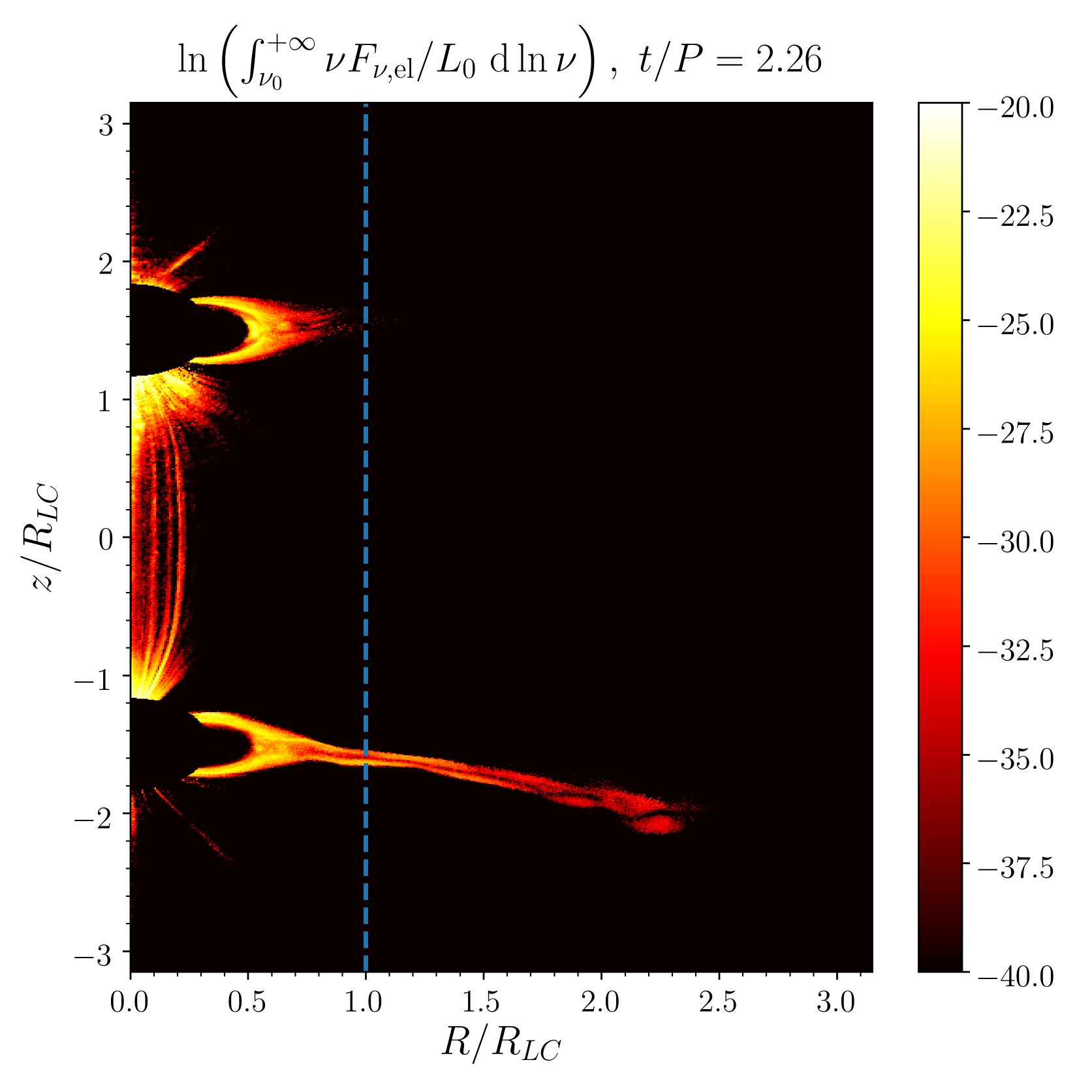}\label{fig:rad_el_same_anti}} 
	\sidesubfloat[]{\includegraphics[width=8.0cm]{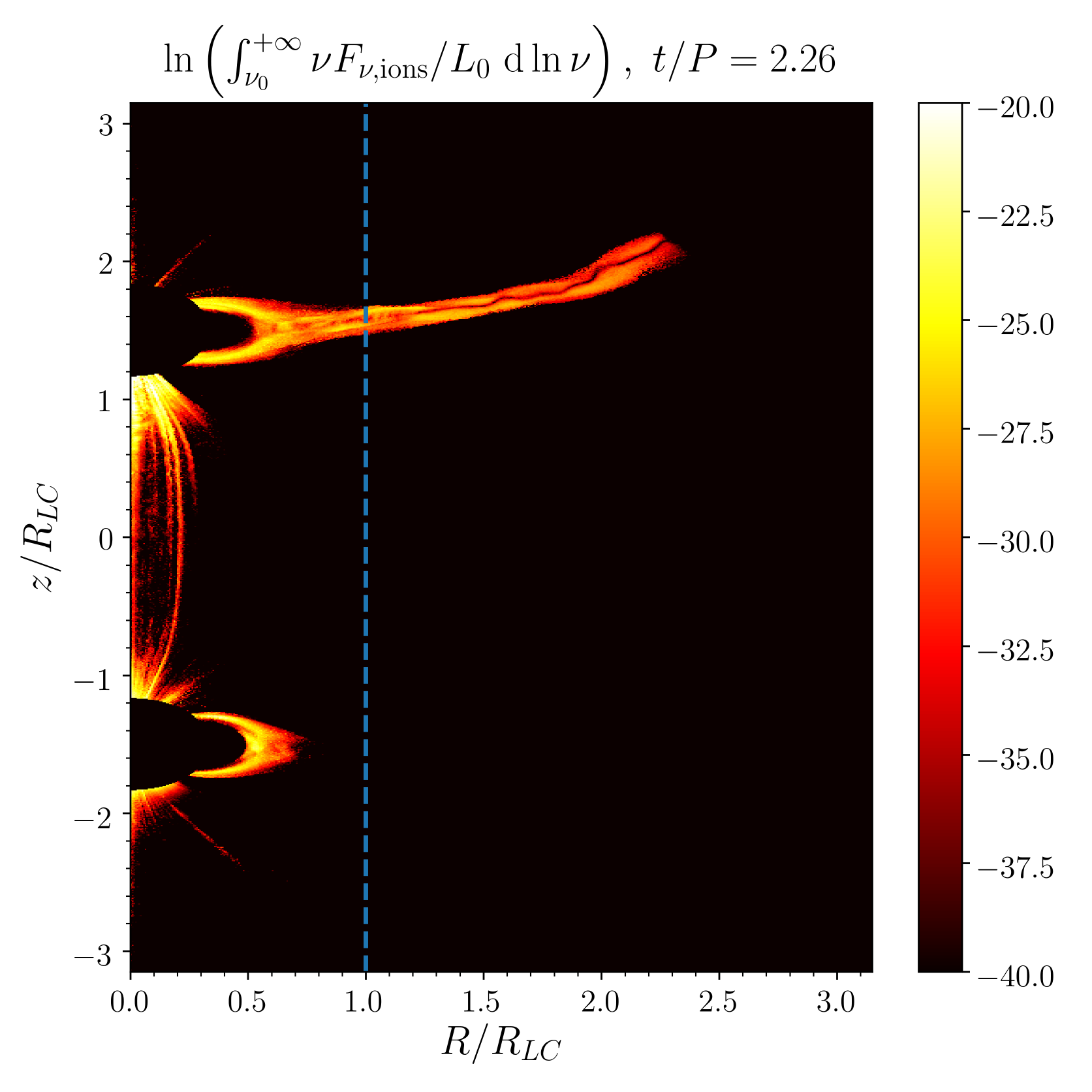}\label{fig:rad_ions_same_anti}} 
	}
	\caption{High-energy radiation maps, as defined in Sec. \ref{sec:radiation}, for the parallel configuration (electrons \textbf{(a)} and positrons \textbf{(b)}) and anti-parallel configuration (electrons \textbf{(c)} and positrons \textbf{(d)}).}
\end{figure*}

The high-energy radiation maps for electrons and positrons are shown in Figs. \ref{fig:rad_el_same_aligned}, \ref{fig:rad_ions_same_aligned} for the parallel configuration, Figs. \ref{fig:rad_el_same_anti} and \ref{fig:rad_ions_same_anti} for the anti-parallel configuration. As mentioned earlier, particles too close to the stellar surface are spurious and should not contribute to the radiation diagnostic. The anti-parallel maps show that at large separation, high-energy radiation is mainly due to the current sheets proper to each pulsar. Besides, in accordance with the diagnostic of Fig. \ref{fig:alpha_same_anti}, particle acceleration is enhanced at the inward poles, with respect to the outward polar cap emission. In the parallel configuration, the inter-pulsar current sheet radiates predominantly, as expected since the Y-point lies within the light cylinder where the field is stronger. From the radiation data, one can infer how much of the spindown was channeled to electromagnetic radiation, by computing the total radiative efficiency in the total simulation domain $\mathcal{V}$ (excluding a thin shell of width $0.05 \, R_0$ at the stellar surface):
\begin{equation} \label{eq:efficiency}
 \eta_\mathrm{rad} = \dfrac{1}{L_0} \int_0^{+\infty} \mathrm{d} \ln \, \nu \int_{\mathcal{V}} \mathrm{d}V \ \nu (F_{\nu,\mathrm{el}} + F_{\nu,\mathrm{ions}}) .
\end{equation}
In this study, the radiative efficiency of an isolated pulsar is $\eta_\mathrm{rad} = 2.7 \%$. The emitted radiation in the steady state two-pulsar setup is much more intense than in the isolated pulsar case. At $a=3 \, R_{LC}$, we find $\eta_\parallel=22 \%$ in the parallel configuration and $\eta_{\centernot\parallel}=23 \%$ in the anti-parallel configuration. At the closer separation $a=1.25 \, R_{LC}$, the efficiencies are $\eta_\parallel=29 \%$ and $\eta_{\centernot\parallel} = 81 \%$ respectively.
\smallskip

To sum up, even though the reconnection mechanism is more efficient at dissipating electromagnetic energy than the DC mechanism, the amplification of $B_\varphi$ in the anti-parallel configuration compensates for its relative inefficiency. The amount of \emph{dissipated} energy is similar in both cases (around $1.0 \, L_0$ at $a=1.25 \, R_{LC}$, inferred from Tab. \ref{tab:poynting}). Eventually, the radiative efficiency increases faster with decreasing separation in the anti-parallel configuration. This can be explained by the low densities achieved in the inter-pulsar reconnection layer, with respect to the dense zone between the stars. Although particles experience greater accelerations in the parallel configuration, this affects a smaller number of particles, so that \emph{in fine} more energy is converted into particle kinetic energy in the anti-parallel configuration.

\subsection{The inspiral phase} \label{sec:inspiral}

\begin{figure}[ht!]
    \resizebox{\hsize}{!}{\includegraphics{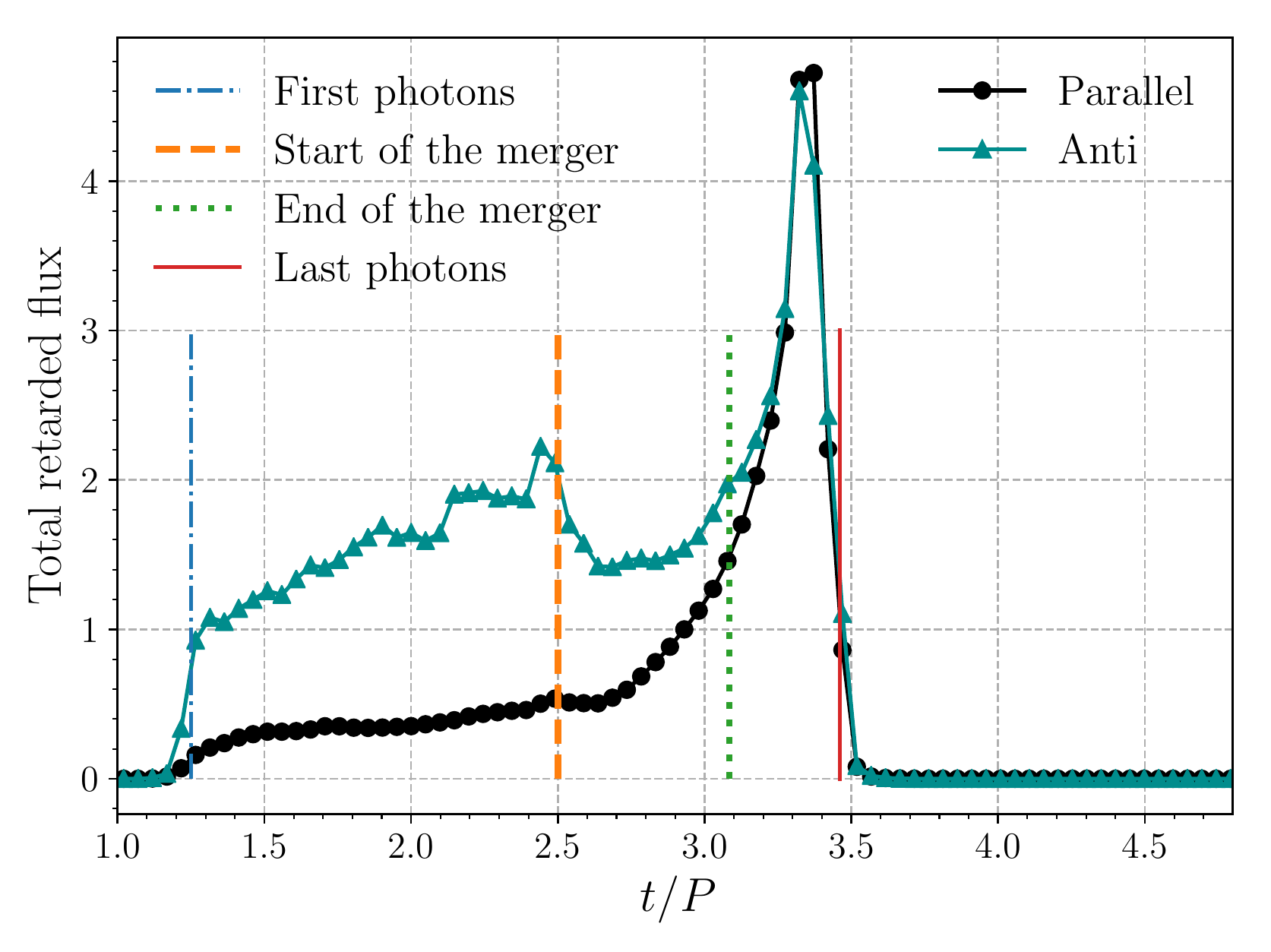}}
    \caption{Lightcurves in the parallel (black circles) and anti-parallel (blue triangles) configurations, normalized by the isolated pulsar spindown power $L_0$. The diagnostic starts at the dash-dotted line. The merger starts at the dashed line, until the two stars touch at the dotted line. Photons are still collected by the observer until the solid line is reached. The time is normalized by the pulsar's spin period.}
    \label{fig:lightcurve}
\end{figure}

The lightcurves obtained using the procedure described in Sec. \ref{sec:radiation} are presented in Fig. \ref{fig:lightcurve}. In these simulations, the computation of the lightcurves started after $1.25 \, P$ (marked by a dash-dotted line on Fig. \ref{fig:lightcurve}), while the inspiral was initialized at $2.5 \, P$ (marked by a dashed line). Computing the lightcurve after a steady state is reached provides a baseline to evaluate the increase in the measured flux, and makes it possible to check that the routine works correctly by comparing the steady state value to the radiative efficiency.

The two curves are qualitatively different before the inspiral starts. In the parallel simulation, a steady state is reached when the merger is initiated. Conversely, the anti-parallel lightcurve has not yet reached a steady state when we start approaching the stars. The received flux rises steadily. The regular increase in radiated flux is due to the accumulation of toroidal field between the pulsars. This can be confronted with the DC model mentioned earlier. \citet{Lai_merger} noticed that flux tubes twisted beyond $B_\varphi / B_z \gtrsim 1$ break down, because the magnetic pressure exerted by the toroidal field is too large. This occurs if the resistivity of the plasma is too low, and the currents too high. The axisymmetry of the tube is broken by unstable kink modes, which distort the tube. Nonlinear evolution of the tube disrupts it completely. Magnetic energy is then dissipated by reconnection. Thus a quasi-periodic circuit can be expected: after the flux tube breaks, reconnection between the inflated field lines restores the linkage between the two stars and the cycle repeats. However, this phenomenon can only be captured in 3D simulations. The staircase look of the anti-parallel lightcurve possibly results from a discharge phenomenon similar to what was explained in Sec. \ref{sec:magnetospheric}. Nonetheless, the presence of such twist in 2D simulations is promising as to the strength of the outburst, since even more energy should be released by the breaking of the flux tube.

\smallskip 

\begin{figure}[ht!]
    \resizebox{\hsize}{!}{\includegraphics{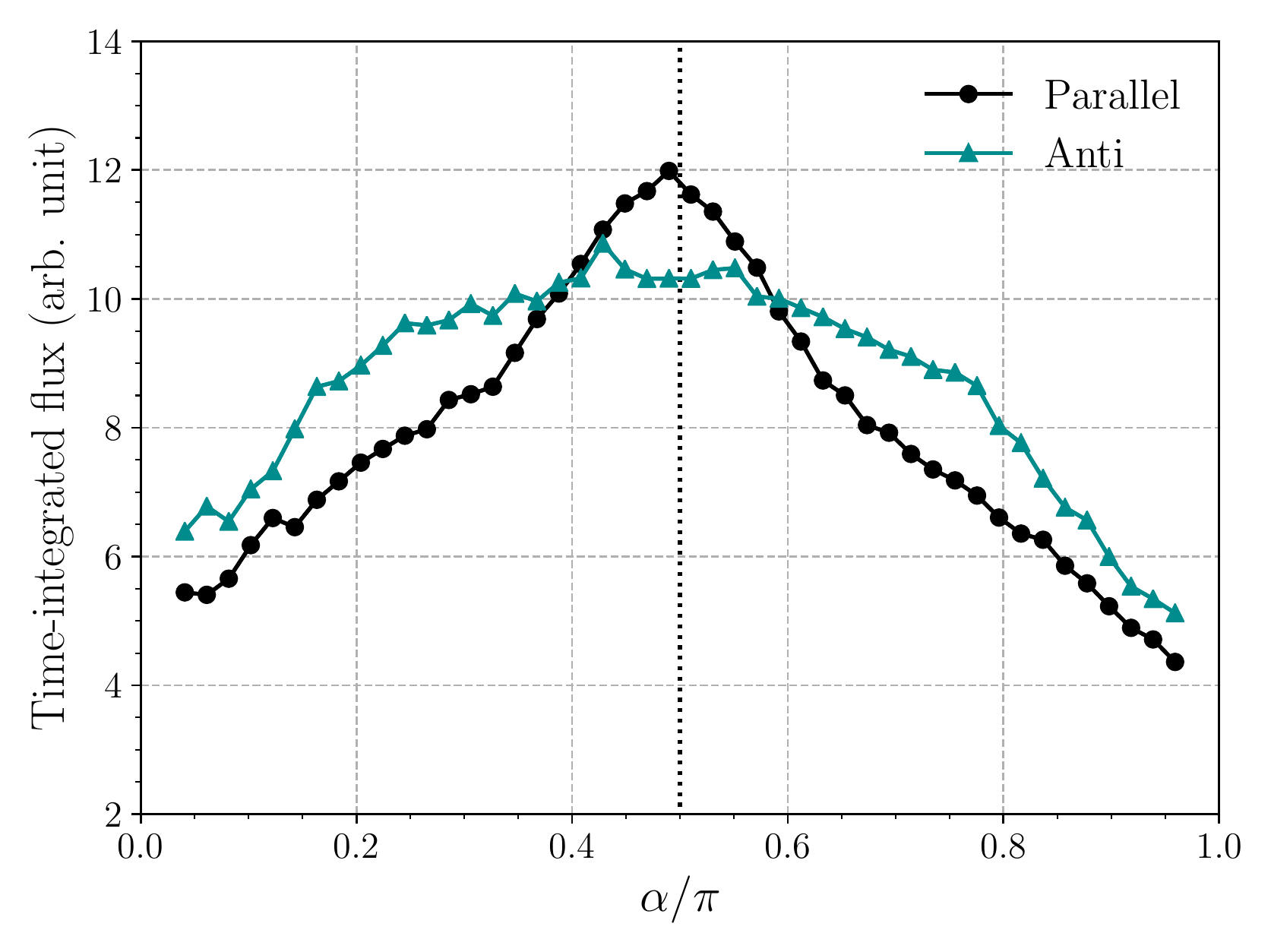}}
    \caption{Angular distribution of the lightcurve signal during the inspiral phase in the two configurations. Time integration ran from the start of the merger until the end of the pulse. The dotted line marks the equator $\alpha_\mathrm{obs}=\pi/2$.}
    \label{fig:lightcurve_angle}
\end{figure}

After the merger starts, both curves eventually meet. The peak of the lightcurve is reached after the end of the merger (marked by the dotted line in Fig. \ref{fig:lightcurve}). The full width at half maximum of the peak is $T_\mathrm{merge} \approx 0.2 \, P$. This can be compared to the final orbital frequency in the GW170817 neutron star merger $f_\mathrm{max}\approx \unit{400}{\hertz}$ \citep{Gravitational_waves_2,Gravitational_waves}. This means that the energy of the magnetospheric outburst is released during the last orbit of the binary. The angular distribution of the pulse is shown in Fig. \ref{fig:lightcurve_angle}. The radiation is mainly emitted in the equatorial plane in both configurations, although the signal is not strongly anisotropic. Consequently, the direction of the line of sight is not critical regarding observability. The anti-parallel angular distribution is broader than the parallel one, because the emission mechanism occurring before the merger is different. In contrast to magnetic reconnection, which mainly accelerates particles near the equatorial plane, in the anti-parallel configuration particles are accelerated when heading from one pulsar to the other, thus pointing towards high latitudes.
\smallskip

The similarity between the two lightcurves in the late phase strongly indicates that close to the merger, the physics that accelerates particles and dissipates energy is the same. In both configurations, there is a \emph{toroidal} current sheet resulting from the discontinuity in the poloidal magnetic field. At high separations its effect is negligible with respect to the other current sheets (the proper pulsar current sheets, or the inter-pulsar radial current sheet in the parallel configuration). However, as the pulsars close up, it becomes predominant in both cases because the poloidal field dominates over the toroidal component. The snapshots of $J_\varphi / c \rho_{GJ}$ right before the pulsars touch show that the toroidal sheet is indeed predominant, and that it has a similar shape in both configurations. At the end of the inspiral, dissipation mainly occurs in the toroidal current sheet through magnetic reconnection, which converts a great amount of energy, as the fields so close to the stars are huge. Plus, magnetic reconnection is forced as the pulsars close up, which increases the reconnection rate. 

\begin{figure}[ht!]
    \resizebox{\hsize}{!}{\includegraphics{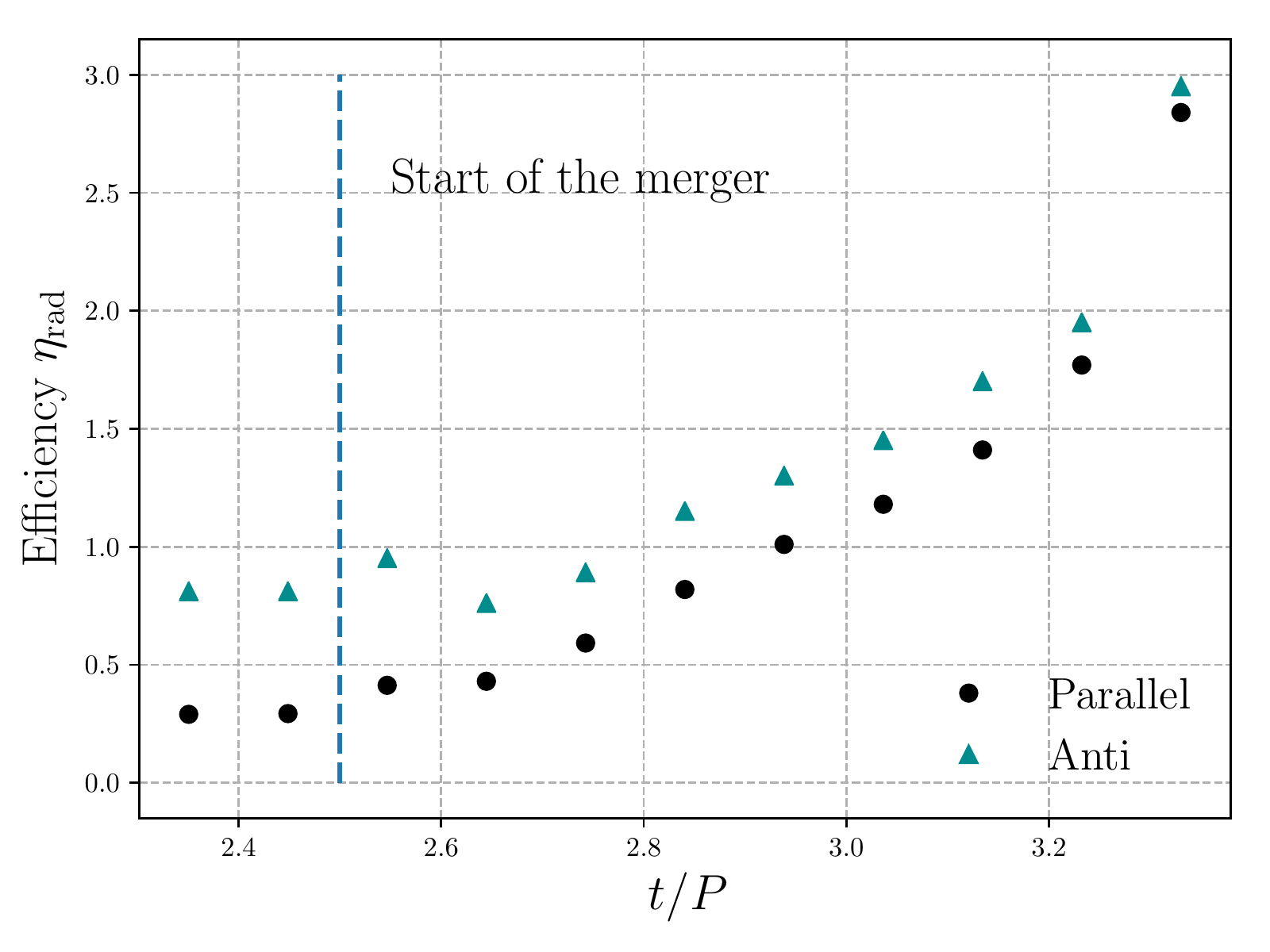}}
    \caption{Total instantaneous radiative efficiency $\eta_\mathrm{rad}$, given by Eq. \eqref{eq:efficiency}, plotted as a function of time in both configurations. The beginning of the merger is marked by the dashed line. The time is normalized by the pulsars' spin period.}
    \label{fig:efficiency}
\end{figure}

At the end of the day, the pulsar inspiral leads to a great increase in radiated power. The instantaneous radiative efficiency (displayed in Fig. \ref{fig:efficiency}) shows an increase by a factor $\sim 10$ of the bolometric luminosity for both configurations between the beginning and the end of the inspiral phase. This amounts to an increase by a factor $\sim 100$ with respect to the case of two quasi-isolated pulsars, with $a\gg R_{LC}$. Even so, the dissipated power is well below the previous theoretical expectations from the DC model. Including the orbital motion in 3D simulations would probably increase the simulated dissipated energy, as the relative motion and thus the electromotive force would be greater. However, these theoretical estimates concern the total output energy, yet only the radiated energy will leave an observable signature.

\subsection{Asymmetric simulations} \label{sec:asymmetric}

\begin{figure*}[ht!]
    \centering
    \captionsetup[subfigure]{position=top, labelfont=bf,textfont=normalfont,singlelinecheck=off,justification=raggedright}
	\makebox[\textwidth][c]{
	\sidesubfloat[]{\includegraphics[width=9.0cm]{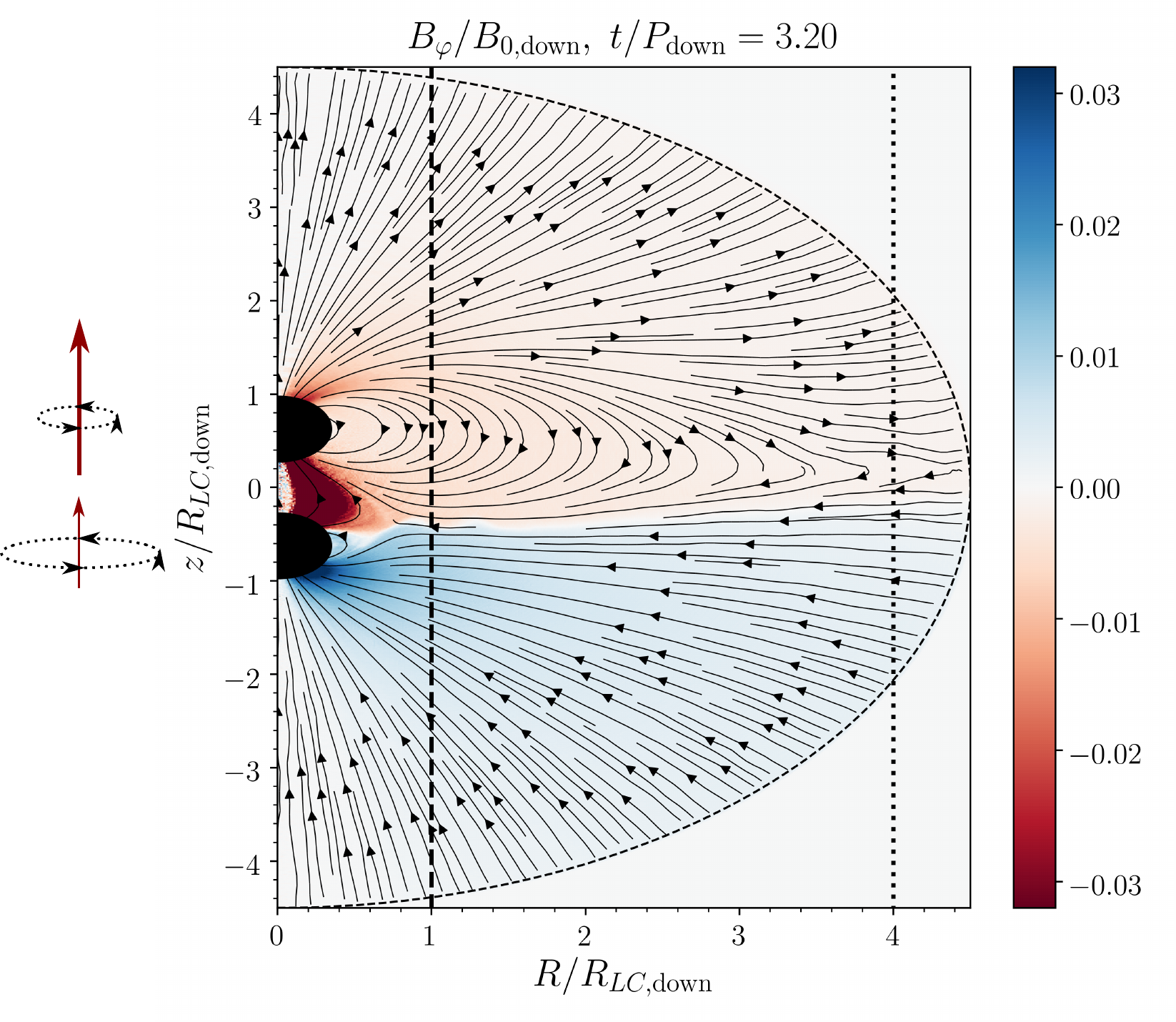}\label{fig:Bth_asy_same}} 
	\sidesubfloat[]{\includegraphics[width=9.0cm]{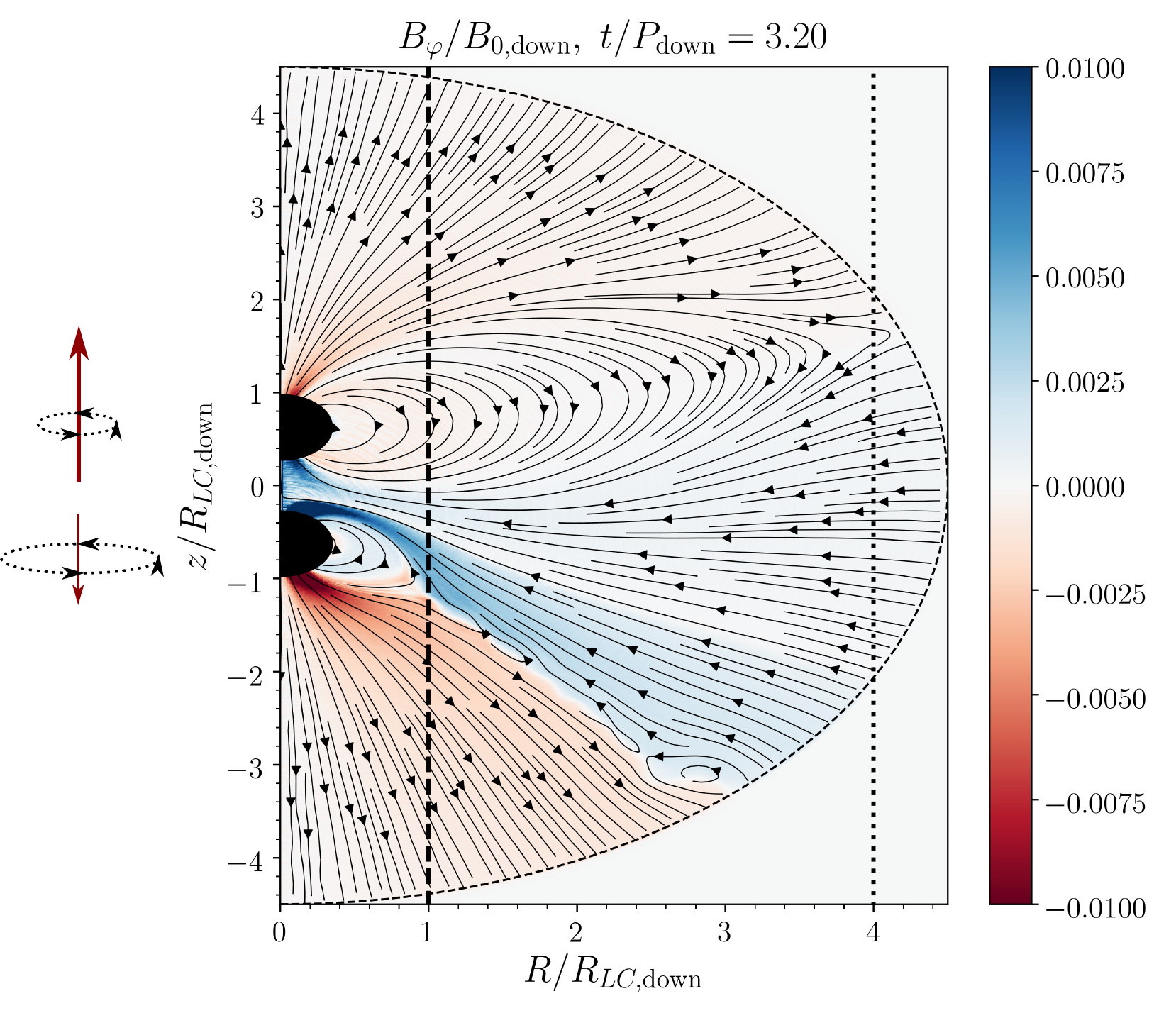}\label{fig:Bth_asy_opposed}} 
	}
	\caption{Normalized toroidal magnetic field $B_\varphi /B_{0,\mathrm{down}}$ for asymmetric simulation with $\Omega_\mathrm{down} / \Omega_\mathrm{up} = 4$ and $B_{0,\mathrm{down}}/B_{0,\mathrm{up}}=0.25$, in the configuration with identical spins and aligned \textbf{(a)} or anti-aligned \textbf{(b)} magnetic moments. The simulation has reached a steady state. The young (resp. old) pulsar is at the top (resp. bottom). The poloidal magnetic field lines are depicted as black solid lines. The black dashed (resp. dotted) line denotes the light cylinder of the bottom (resp. top) pulsar. The stellar separation is fixed at $a=1.25 R_{LC}$.}
\end{figure*}

So far we have focused on symmetric binaries. Actual binary pulsars are likely to be asymmetric \citep{Tauris,Piro}, comprising a recycled pulsar spinning rapidly and a younger, slower pulsar with a stronger magnetic field. This was confirmed by the discovery of the double pulsar J0737-3039A/B \citep{Kramer}, for which $\Omega_1 / \Omega_2 \approx 10^2$ and $B_1 / B_2 \approx 5\times 10^{-3}$. However, such ratios are far too extreme for our setup. The light cylinders of both pulsars must be included in the simulation domain, otherwise the closed magnetic field lines of the slow pulsar interact with the outer damping layer. In this case the closed corotating zone is artificially reduced, which induces spurious particle acceleration. Instead, we performed simulations with $\Omega_1 / \Omega_2 = 4$ and $B_1/B_2 = 0.25$, and increased the size of the box to include the whole magnetosphere of the slow pulsar. The bottom pulsar is the old recycled pulsar; it has the same spin and magnetic strength than in the symmetric simulations, whereas the top pulsar has a stronger field and slower spin. 
\smallskip

A map of the normalized toroidal field in the configuration with aligned spins and magnetic moments is shown in Fig. \ref{fig:Bth_asy_same}. In contrast to symmetric simulations, the relative orientations of the spins matter little. The parallel and anti-parallel configurations are similar, since the toroidal field of the fast recycled pulsar dominates anyway. Strongly twisted magnetic field lines connect the two pulsars, like in the symmetric anti-parallel setup. The closed zone of the weak pulsar is compressed because of the influence of the strong one above it. The inter-pulsar current sheet present in the symmetric parallel configuration (see Fig. \ref{fig:Bth_same_aligned}) is missing in Fig. \ref{fig:Bth_asy_same}. Snapshots of the toroidal field at early times show that these two current sheets merge, as the toroidal field of the fast pulsar takes over. The resulting current sheet is deflected upwards in this parallel configuration, but downwards if they are anti-parallel. This is because the compensation of the top pulsar toroidal field makes the magnetic pressure initially smaller above the down pulsar than below.
\smallskip

In this situation it is interesting to study also the case of two opposed magnetic moments. As shown in Fig. \ref{fig:Bth_asy_opposed}, the situation is quite different. No field line can connect the two stars, so the influence of the strong pulsar on the weak one can only be indirect. Plus, no current can flow from one pulsar to the other. As a result, the relative orientation of the spins has rigorously no impact on the magnetic topology, the polarities of the toroidal field and currents due to the down pulsar are simply reversed. At $t=0$, the magnetosphere of the weak pulsar is confined within the strong field of the other one. Its magnetic field lines then open up and extend to infinity as rotation is imposed. The current sheet of the weak pulsar is strongly deflected. Remarkably, at high separations, this configuration radiates no more than the isolated pulsar. However, after the merger starts, the radiated flux strongly increases by a factor $\sim 100$, until it reaches usual outburst values (see Tab. \ref{tab:results}). 
\smallskip

We identify multiple dissipation mechanisms. Magnetic reconnection occurs at the Y-point of the current sheet, which is pushed closer to the star by the strong pulsar. In the case of aligned magnetic moments, strong pair creation and particle acceleration occur between the pulsars, as in the symmetric anti-parallel configuration. The higher outburst values in asymmetric simulations can be understood in this framework (see Tab. \ref{tab:results}). Asymmetric simulations exhibit an intense magnetic field, which compresses the old pulsar's magnetosphere. The Y-point at the foot of the recycled pulsar's proper current sheet lies within its light cylinder. Reconnection is therefore more efficient than in symmetric simulations. We also think that in contrast to symmetric simulations, emission from the north polar cap of the strong pulsar may be essential. High-energy emission in this area has very contrasted stripes, which indicates a violent pair creation cascade. The presence of the fast pulsar below implies strong currents must flow from the south polar cap of the strong pulsar. Therefore strong currents must also flow from its north polar cap, and pair creation is triggered to supply them. Then the huge surface parallel electric field accelerates particles. This last mechanism was negligible in symmetric simulations. However, the radiation diagnostics in the asymmetric simulations must be taken with care. Since the magnetic field of the top pulsar is increased in this setup, the plasma quantities are less resolved, and the output data is not fully reliable close to the pulsars. The exact acceleration processes are thus harder to infer.


\section{Discussion, observational prospects}

We have performed global PIC simulations of a binary pulsar, with various magnetic moments and spins. Our strongest conclusion is that the total radiated power increases by two orders of magnitude during the whole inspiral. The bolometric luminosity can reach up to ten times the spindown power of an isolated pulsar. The lightcurve presents a peak whose width is roughly $0.2 \, P$, which approximately corresponds to the last orbit of the double pulsar. The shape of the peak depends very little on the relative orientation of the spins or magnetic moments. The radiative power is concentrated within the equatorial regions but presents a weak anisotropy. We find that magnetic reconnection is a key ingredient in for particle acceleration and the emission of high-energy synchrotron radiation in all the configurations explored here. In the symmetric configuration, dissipation occurs mainly because of an inter-pulsar current sheet forming at the interface between the two pulsar magnetospheres. As a result, electromagnetic energy is released even in the case where the two pulsars spin synchronously. Before the last stage of the merger, a significant amount of energy is also dissipated between the two stars in the symmetric anti-parallel configuration. Besides, pair creation at the outward poles is amplified with respect to the isolated pulsar configuration. In particular, in asymmetric simulations, pair creation in the vicinity of the  slow magnetized pulsar is revived by the presence of the fast-spinning pulsar.
\smallskip

\begin{table}[h!]
\centering
   \captionsetup{width=0.9\textwidth}
\begin{tabular}{DEE}
\hline\hline
 & Symmetric & Asymmetric\\
\hline
  Same parallel &  $4.9 \, L_0$ &  $8.0 \, L_{0,\mathrm{down}}$ \\
  Same anti-parallel &  $5.0 \, L_0$ &  $10 \, L_{0,\mathrm{down}}$ \\
  Opposed parallel &  $1.9 \, L_0$ &  $4.4 L_{0,\mathrm{down}}$ \\
\hline
\end{tabular}
\caption{Maximum total radiated power for different configurations. ``Same'' (resp. ``Opposed'') denotes aligned (resp. anti-aligned) magnetic moments, whereas ``parallel'' (resp. ``anti-parallel'') denotes aligned (resp. anti-aligned) spins. $L_0$ is the spindown power of an isolated pulsar. $L_{0,\mathrm{down}}$ is the spindown of the bottom (recycled) pulsar if it were isolated. In all our asymmetric runs, the bottom pulsar was assigned the same magnetic strength and rotation period as in the symmetric simulations.}
\label{tab:results}
\end{table}

A natural extension of this work would be to perform 3D PIC simulations of a binary aligned pulsar, and more generally with arbitrary inclinations. This would have several benefits. First, we would be able to capture plasma instabilities that can only develop in 3D. For instance, the symmetric anti-parallel configuration displays a highly twisted flux tube, which should be unstable in a 3D setup. This could be another source of energy dissipation. Second, a 3D setup would allows us to simulate more realistic configurations. But more importantly, we would be able to take orbital motion into account. Since the orbital frequency peaks at the merger, it might lead to an increased electromotive force, stronger currents and more dissipation. The luminosity of the precursor predicted here should therefore be seen as a lower limit. Notwithstanding this, for the first time we are able to self-consistently solve the magnetospheric interaction between two coalescing pulsars. This simplified setup allows us to handily elucidate the physical mechanisms at play.
\smallskip

Assuming a binary pulsar with a powerful Crab-like pulsar \citep{Crab_nebula}, we predict a luminosity for the electromagnetic precursor of its coalescence of $\sim 10^{38}$ erg/s. In the optimistic case where most of this energy is emitted in the \emph {Fermi}/GBM range, this sets a distance upper limit of $\lesssim 4$ Mpc for detection. Thus, only the local group galaxies could be probed this way. For comparison, the neutron star merger event GRB 170817A, which occurred at a distance of $40$ Mpc, released an amount of $\sim 10^{46}$ erg/s in the $\gamma$ range \citep{Gravitational_waves_2}. The situation is similarly grim in X-ray, but radio emission could be of better use. Some radio detectors such as MeerKAT, or the SKA array in a near future, map the whole sky in real time with high sensitivity. There are some caveats however: the radio luminosity of most pulsars is only a small faction of the spindown power (usually around $10^{-6}$). Dispersion by the interstellar medium is also likely to delay the arrival of the radio burst, which cannot arrive ahead of the merger. Besides, radio emission is coherent and does not directly originate from particle acceleration, but may rather be related to pair creation \citep{Philippov_RG}. Although we observe a strong pair creation cascade at the polar caps, especially during the inspiral phase, we are not able to compute the radio efficiency of the binary. Still, radio detection is a promising candidate for the prospects of neutron star merger observability. In particular, Fast Radio Bursts have duration and energy range consistent with pulsar coalescences, and could be counterparts to non-repeating events \citep{FRB}.

\begin{acknowledgements}
We would like to thank the referee Richard Henriksen for valuable comments on the manuscript. We acknowledge the support from CNES and the Universit\'e Grenoble Alpes (IDEX-IRS grant). This work was granted access to the HPC resources of CINES on Occigen under the allocation A0030407669 made by GENCI. 

\end{acknowledgements}


\bibliographystyle{aa}
\bibliography{biblio}

\begin{thebibliography}{45}
\expandafter\ifx\csname natexlab\endcsname\relax\def\natexlab#1{#1}\fi

\bibitem[{{Abbott} {et~al.}(2017{\natexlab{a}}){Abbott}, {Abbott}, {Abbott},
  {Acernese}, {Ackley}, {Adams}, {Adams}, {Addesso}, {Adhikari}, {Adya}, \&
  et~al.}]{Gravitational_waves_2}
{Abbott}, B.~P., {Abbott}, R., {Abbott}, T.~D., {et~al.} 2017{\natexlab{a}},
  \apjl, 848, L13

\bibitem[{{Abbott} {et~al.}(2017{\natexlab{b}}){Abbott}, {Abbott}, {Abbott},
  {Acernese}, {Ackley}, {Adams}, {Adams}, {Addesso}, {Adhikari}, {Adya}, \&
  et~al.}]{Gravitational_waves}
{Abbott}, B.~P., {Abbott}, R., {Abbott}, T.~D., {et~al.} 2017{\natexlab{b}},
  Physical Review Letters, 119, 161101

\bibitem[{{Beloborodov}(2008)}]{Beloborodov}
{Beloborodov}, A.~M. 2008, \apjl, 683, L41

\bibitem[{{Belyaev}(2015)}]{2015MNRAS.449.2759B}
{Belyaev}, M.~A. 2015, \mnras, 449, 2759

\bibitem[{{Beskin} {et~al.}(1993){Beskin}, {Gurevich}, \& {Istomin}}]{Beskin}
{Beskin}, V.~S., {Gurevich}, A.~V., \& {Istomin}, Y.~N. 1993, {Physics of the
  pulsar magnetosphere}

\bibitem[{Birdsall \& Langdon(2005)}]{Birdsall}
Birdsall, C.~K. \& Langdon, A.~B. 2005, Plasma Physics via Computer Simulation,
  1st edn. (Taylor \& Francis)

\bibitem[{{Blumenthal} \& {Gould}(1970)}]{Blumenthal}
{Blumenthal}, G.~R. \& {Gould}, R.~J. 1970, Reviews of Modern Physics, 42, 237

\bibitem[{{Brambilla} {et~al.}(2018){Brambilla}, {Kalapotharakos}, {Timokhin},
  {Harding}, \& {Kazanas}}]{2018ApJ...858...81B}
{Brambilla}, G., {Kalapotharakos}, C., {Timokhin}, A.~N., {Harding}, A.~K., \&
  {Kazanas}, D. 2018, \apj, 858, 81

\bibitem[{{B{\"u}hler} \& {Blandford}(2014)}]{Crab_nebula}
{B{\"u}hler}, R. \& {Blandford}, R. 2014, Reports on Progress in Physics, 77,
  066901

\bibitem[{Carroll(2003)}]{RG}
Carroll, S.~M. 2003, Spacetime and Geometry: An Introduction to General
  Relativity, 1st edn. (Pearson)

\bibitem[{{Cerutti} \& {Beloborodov}(2017)}]{Cerutti_review}
{Cerutti}, B. \& {Beloborodov}, A.~M. 2017, \ssr, 207, 111

\bibitem[{{Cerutti} {et~al.}(2016{\natexlab{a}}){Cerutti}, {Mortier}, \&
  {Philippov}}]{2016MNRAS.463L..89C}
{Cerutti}, B., {Mortier}, J., \& {Philippov}, A.~A. 2016{\natexlab{a}}, \mnras,
  463, L89

\bibitem[{{Cerutti} {et~al.}(2015){Cerutti}, {Philippov}, {Parfrey}, \&
  {Spitkovsky}}]{Cerutti_main}
{Cerutti}, B., {Philippov}, A., {Parfrey}, K., \& {Spitkovsky}, A. 2015,
  \mnras, 448, 606

\bibitem[{{Cerutti} \& {Philippov}(2017)}]{2017A&A...607A.134C}
{Cerutti}, B. \& {Philippov}, A.~A. 2017, \aap, 607, A134

\bibitem[{{Cerutti} {et~al.}(2016{\natexlab{b}}){Cerutti}, {Philippov}, \&
  {Spitkovsky}}]{Cerutti_light_curves}
{Cerutti}, B., {Philippov}, A.~A., \& {Spitkovsky}, A. 2016{\natexlab{b}},
  \mnras, 457, 2401

\bibitem[{{Cerutti} {et~al.}(2013){Cerutti}, {Werner}, {Uzdensky}, \&
  {Begelman}}]{Zeltron}
{Cerutti}, B., {Werner}, G.~R., {Uzdensky}, D.~A., \& {Begelman}, M.~C. 2013,
  \apj, 770, 147

\bibitem[{{Chen} \& {Beloborodov}(2014)}]{2014ApJ...795L..22C}
{Chen}, A.~Y. \& {Beloborodov}, A.~M. 2014, \apjl, 795, L22

\bibitem[{{Dionysopoulou} {et~al.}(2015){Dionysopoulou}, {Alic}, \&
  {Rezzolla}}]{Dionysopoulou}
{Dionysopoulou}, K., {Alic}, D., \& {Rezzolla}, L. 2015, \prd, 92, 084064

\bibitem[{{Goldreich} \& {Julian}(1969)}]{Goldreich_Julian}
{Goldreich}, P. \& {Julian}, W.~H. 1969, \apj, 157, 869

\bibitem[{{Goldreich} \& {Lynden-Bell}(1969)}]{Lynden_bell}
{Goldreich}, P. \& {Lynden-Bell}, D. 1969, \apj, 156, 59

\bibitem[{{Hansen} \& {Lyutikov}(2001)}]{Lyutikov}
{Hansen}, B.~M.~S. \& {Lyutikov}, M. 2001, \mnras, 322, 695

\bibitem[{{Henriksen} \& {Rayburn}(1971)}]{Henriksen}
{Henriksen}, R.~N. \& {Rayburn}, D.~R. 1971, \mnras, 152, 323

\bibitem[{Jackson(1998)}]{Jackson}
Jackson, J.~D. 1998, Classical Electrodynamics, 3rd edn. (Wiley)

\bibitem[{{Kalapotharakos} {et~al.}(2018){Kalapotharakos}, {Brambilla},
  {Timokhin}, {Harding}, \& {Kazanas}}]{2018ApJ...857...44K}
{Kalapotharakos}, C., {Brambilla}, G., {Timokhin}, A., {Harding}, A.~K., \&
  {Kazanas}, D. 2018, \apj, 857, 44

\bibitem[{{Kelner} {et~al.}(2015){Kelner}, {Prosekin}, \& {Aharonian}}]{Kelner}
{Kelner}, S.~R., {Prosekin}, A.~Y., \& {Aharonian}, F.~A. 2015, \aj, 149, 33

\bibitem[{{Kramer} \& {Stairs}(2008)}]{Kramer}
{Kramer}, M. \& {Stairs}, I.~H. 2008, \araa, 46, 541

\bibitem[{{Lai}(2012)}]{Lai_merger}
{Lai}, D. 2012, \apjl, 757, L3

\bibitem[{{Metzger} \& {Zivancev}(2016)}]{Metzger}
{Metzger}, B.~D. \& {Zivancev}, C. 2016, \mnras, 461, 4435

\bibitem[{{Michel}(1973{\natexlab{a}})}]{Michel_1}
{Michel}, F.~C. 1973{\natexlab{a}}, \apj, 180, 207

\bibitem[{{Michel}(1973{\natexlab{b}})}]{Michel_2}
{Michel}, F.~C. 1973{\natexlab{b}}, \apjl, 180, L133

\bibitem[{{Palenzuela} {et~al.}(2013){Palenzuela}, {Lehner}, {Liebling},
  {Ponce}, {Anderson}, {Neilsen}, \& {Motl}}]{Palenzuela}
{Palenzuela}, C., {Lehner}, L., {Liebling}, S.~L., {et~al.} 2013, \prd, 88,
  043011

\bibitem[{{Parfrey} {et~al.}(2012){Parfrey}, {Beloborodov}, \&
  {Hui}}]{Parfrey_alpha_plot}
{Parfrey}, K., {Beloborodov}, A.~M., \& {Hui}, L. 2012, \mnras, 423, 1416

\bibitem[{{Pen}(2018)}]{FRB}
{Pen}, U.-L. 2018, ArXiv e-prints [\eprint[arXiv]{1811.00605}]

\bibitem[{{Philippov} {et~al.}(2014){Philippov}, {Tchekhovskoy}, \&
  {Li}}]{Philippov_obliquity}
{Philippov}, A., {Tchekhovskoy}, A., \& {Li}, J.~G. 2014, \mnras, 441, 1879

\bibitem[{{Philippov} {et~al.}(2015{\natexlab{a}}){Philippov}, {Cerutti},
  {Tchekhovskoy}, \& {Spitkovsky}}]{Philippov_RG}
{Philippov}, A.~A., {Cerutti}, B., {Tchekhovskoy}, A., \& {Spitkovsky}, A.
  2015{\natexlab{a}}, \apjl, 815, L19

\bibitem[{{Philippov} \& {Spitkovsky}(2014)}]{2014ApJ...785L..33P}
{Philippov}, A.~A. \& {Spitkovsky}, A. 2014, \apjl, 785, L33

\bibitem[{{Philippov} \& {Spitkovsky}(2018)}]{2018ApJ...855...94P}
{Philippov}, A.~A. \& {Spitkovsky}, A. 2018, \apj, 855, 94

\bibitem[{{Philippov} {et~al.}(2015{\natexlab{b}}){Philippov}, {Spitkovsky}, \&
  {Cerutti}}]{2015ApJ...801L..19P}
{Philippov}, A.~A., {Spitkovsky}, A., \& {Cerutti}, B. 2015{\natexlab{b}},
  \apjl, 801, L19

\bibitem[{{Piro}(2012)}]{Piro}
{Piro}, A.~L. 2012, \apj, 755, 80

\bibitem[{{Sironi} \& {Cerutti}(2017)}]{Pic_review_cerutti}
{Sironi}, L. \& {Cerutti}, B. 2017, in Astrophysics and Space Science Library,
  Vol. 446, Modelling Pulsar Wind Nebulae, ed. D.~F. {Torres}, 247

\bibitem[{{Tauris} {et~al.}(2017){Tauris}, {Kramer}, {Freire}, {Wex}, {Janka},
  {Langer}, {Podsiadlowski}, {Bozzo}, {Chaty}, {Kruckow}, {van den Heuvel},
  {Antoniadis}, {Breton}, \& {Champion}}]{Tauris}
{Tauris}, T.~M., {Kramer}, M., {Freire}, P.~C.~C., {et~al.} 2017, \apj, 846,
  170

\bibitem[{{Tauris} \& {Manchester}(1998)}]{Tauris_obliquity}
{Tauris}, T.~M. \& {Manchester}, R.~N. 1998, \mnras, 298, 625

\bibitem[{{Vietri}(1996)}]{1996ApJ...471L..95V}
{Vietri}, M. 1996, \apjl, 471, L95

\bibitem[{{Yee}(1966)}]{1966ITAP...14..302Y}
{Yee}, K. 1966, IEEE Transactions on Antennas and Propagation, 14, 302

\bibitem[{{Young} {et~al.}(2010){Young}, {Chan}, {Burman}, \&
  {Blair}}]{Young_obliquity}
{Young}, M.~D.~T., {Chan}, L.~S., {Burman}, R.~R., \& {Blair}, D.~G. 2010,
  \mnras, 402, 1317

\end{thebibliography}

\end{document}